\RequirePackage{lineno} 
\documentclass[12pt]{article}
\usepackage{graphicx,psfrag,epsf,adjustbox}
\usepackage{enumerate}
\usepackage[english]{babel}
\usepackage{amssymb,amsmath,amsthm,amsfonts,mathtools}
\usepackage{caption,subfig}
\usepackage{bm}
\usepackage{color}
\usepackage{lineno}
\usepackage{url}
\usepackage{natbib,hyperref}
\usepackage{multirow}

\newcommand{\data}{\bm{x}}
\newcommand{\dataObs}{\bm{x^0}}
\newcommand{\datasim}{\bm{x^{\mbox{sim}}}}
\newcommand{\distance}{d}
\newcommand{\Model}{\mathcal{M}}
\DeclareMathOperator*{\argmin}{arg\,min}
\newcommand{\parameter}{\bm{\theta}}
\newcommand{\parametertheo}{\bm{\phi}}
\newcommand{\estparametertheo}{\bm{\hat{\phi}}}
\newcommand{\prior}{\pi}

\newcommand{\threshold}{\gamma}

\newcommand{\obsstarttime}{t_0}
\newcommand{\obsendtime}{T}

\newcommand{\lossfunc}{\mathcal{L}}
\newcommand{\summaryfunc}{\mathcal{F}}
\newcommand{\distdata}{d}

\newcommand{\eucldist}{d_{E}}

\newcommand{\blind}{0}
\begin{document}

\if0\blind
{

\title{\bf Bayesian Calibration of Force-fields from Experimental Data: TIP4P 
Water} 

\author{Ritabrata Dutta$^1$\thanks{Corresponding author: duttar@usi.ch},
Zacharias Faidon Brotzakis$^{1,2}$,\\
Antonietta Mira$^{1,3}$\\ {\em \small $^1$Institute of Computational Science, Universit\`a della Svizzera italiana, Switzerland}\\
 {\em \small $^2$ Department of Chemistry and Applied Bioscience, ETH Z{\"u}rich}\\
{\em \small$^3$Department of Science and High Technology, Universit\`a degli Studi dell'Insubria, Italy}\\  }

  \maketitle } \fi

\if1\blind
{
  \bigskip
  \bigskip
  \bigskip
  \begin{center}
    {\LARGE\bf Title}
\end{center}
  \medskip
} \fi

\bigskip
Molecular dynamics (MD) simulations give access to equilibrium structures and dynamic properties given an ergodic sampling and an accurate force-field.
The force-field parameters are calibrated to reproduce properties measured by experiments or simulations. The main contribution of this paper is an approximate Bayesian framework for the calibration and uncertainty quantification
of the force-field parameters, without assuming parameter uncertainty to be Gaussian. 
To this aim, since the likelihood function of the MD simulation models are intractable in absence of Gaussianity assumption, we use a likelihood-free inference scheme known as approximate Bayesian computation (ABC) and propose an adaptive population Monte Carlo ABC algorithm, which is illustrated to converge faster and scales better than previously used ABCsubsim algorithm for calibration of force-field of a helium system. The second contribution is the adaptation of ABC algorithms for High Performance Computing to MD simulation within the Python ecosystem \emph{ABCpy}. This adaptation includes a novel use of dynamic allocation scheme for MPI. 
We illustrate the performance of the developed methodology to learn posterior distribution and Bayesian estimates of Lennard-Jones force-field parameters of helium and TIP4P system of water implemented
both for \emph{simulated}  and  \emph{experimental datasets} collected
using Neutron and X-ray diffraction. For simulated data, the Bayesian estimate is in close agreement with the true parameter value used to generate the dataset. For experimental as well as for simulated data, the Bayesian posterior distribution shows a  strong correlation pattern between the force-field parameters. 
Providing an estimate of the entire posterior distribution, our methodology also allows us to perform uncertainty
quantification of model prediction.
This research opens up the possibility to rigorously calibrate force-fields from available experimental datasets of any structural and dynamic property.  

\vspace{1ex}\noindent
{\it Keywords: ABCpy, Approximate Bayesian Computation, Bayesian inference, Force-field parameters, Lennard-Jones, TIP4P, High Performance Computing, MPI, Uncertainty quantification.}  

\section{Introduction}
\label{sec:intro}
In the last decades, molecular simulations have become a cornerstone for computing equilibrium and/or dynamic properties of classical many body systems, as well for bridging microscopic with macroscopic observables that could be of both experimental and basic importances  \citep{Allen1989,Frenkel2001,Karplus2014}.
Given a force-field formalism, in this paper we assume that the phase space can be ergodically explored by evolving the Newton's equation of motion under a molecular mechanics force-field of interactions. For this purpose, we consider Molecular Dynamics (MD) simulations, which samples 
the phase space by integrating the deterministic Newtons equations of motion, hence giving access to both dynamic and thermodynamic properties. 
The accuracy of the underlying molecular mechanics force-field used to solve the equations of motion
defines the approximation in the phase space exploration.

Each force-field reproduces specific properties and is indexed by a set of parameters, whose values are unknown (e.g., bonded and non-bonded force-field parameters).
If the parameters of the force-field could be rigorously learned 
with an automated data-driven methodology, we could reparametrize 
the force-field for different experimentally obtained target properties (e.g., radial distribution functions, self-diffusion coefficient, density etc.). 
 There exist different force-field formalisms, constraining ourself only to water we have TIP3P, TIP4P \citep{Jorgensen1983}, TIP4P/2005 \citep{abascal2005general} and TIP5P \citep{mahoney2000five} among  others. This raises the need for calibrations and force-field comparisons in a data-driven manner. Given
 an experimentally observed dataset, the Bayesian inferential framework can address both questions, namely  calibrating the parameters of a given force-field formalism and, given many formalism, choosing one.
 This is achieved, correspondingly, through Bayesian parameter inference and model selection \citep{Marin_2012}. In this paper we only focus on the former.

The underlying uncertainties in MD simulations can be classified in four main 
categories \citep{angelikopoulos2012bayesian} 
(1) \emph{Modeling uncertainty} due to the specific choice of model (e.g., a specific choice of
the functional form for the force-field).
(2) \emph{Parametric uncertainty} due to the unknown values of the 
set of parameters defining the selected model, (e.g., the parameters of the selected force-field function).
(3) \emph{Computational uncertainty} due to the particular computational setup
(e.g., the systematic error due to the box size, fixed number of molecules, finite sampling time 
and solution of Newton's equation by time-integrators), or due to the stochastic components in the computational model (e.g., stochastic thermostats). 
(4) \emph{Measurement uncertainty} due to the experimental or observational error, only occurring while calibration of a force-field is performed based on an experimentally observed dataset.

The use of Bayesian inference for MD simulation has a long history \citep{cailliez2011statistical, angelikopoulos2012bayesian, rizzi2012uncertainty, angelikopoulos2013data, chernatynskiy2013uncertainty, cailliez2014calibration, farrell2015bayesian, sargsyan2015statistical, farrell2017adaptive, messerly2017uncertainty}; for a recent review and a comparative study of different inferential approaches, we direct the readers to \cite{pernot2017critical}.
In most of these works, the different types of uncertainties 
are assumed to be Gaussian in nature. This 
provides us with a simple functional form for likelihood  which in turn allows the use of standard
Bayesian  tools for inference and calibration of force-fields. In the ergodic limit (where the entire state space is explored), and if there is no bias due to under-sampling 
from the Molecular dynamics simulation (e.g. using GROMACS  \citep{Pronk2013} or LAMMPS \citep{plimpton1995fast}), 
approximating the uncertainty 
as Gaussian is theoretically meaningful. But for real life systems, the state space is explored very slowly
due to existence of large free energy barriers, hence a Gaussian model of uncertainty may be a poor 
approximation. Furthermore, the Gaussianity assumption does not hold when one wants to calibrate force-fields based on structural and dynamical properties with non-Gaussian data distribution (e.g. the radial distribution functions and the self-diffusion coefficient)
or the partition function of the NPT-ensemble, 
which is known to be non-Gaussian due to its thermodynamics definition  \citep{Kulakova_2016}.

If we do not assume Gaussianity, 
one of the main difficulty in applying Bayesian inference is the intractability of the likelihood functions of the force-field parameters for an observed dataset. 
In this paper we solve this problem by
approximate Bayesian computation (ABC) \citep{lintusaari2017fundamentals}, a likelihood-free inference scheme recently developed  in the field of statistical science. ABC, has been used for calibration of MD simulation, in \cite{sargsyan2015statistical} under Gaussianity assumption and in \cite{Kulakova_2016} to calibrate parameters of Lennard-Jones (LJ) potential of helium without the assumption of Gaussian uncertainty. \cite{pernot2017critical} concludes that most of the existing algorithms poorly quantify model prediction uncertainty, 
with ABC (under Gaussian assumption of uncertainty) being the most promising one. 
In \cite{Kulakova_2016}, the ABC algorithm, more specifically ABCsubsim   \citep{Chiachio_2014}, is proven to be equivalent to an algorithm that exploits the Gaussian assumption 
for one of their examples,
but is not able to further improve inferential results. 
This may be due to the use of ABC under Gaussian assumption 
of uncertainty \citep{pernot2017critical} or due to the deficiencies in specific ABC algorithms \citep{Kulakova_2016}.  

Following these works, we propose adaptive 
population Monte Carlo ABC (APMCABC) algorithm for ca\-li\-bration of force-fields, without assuming any functional form for the likelihood, and illustrate its speed-up, 
faster convergence and better estimate of the parameter values relative to the ABCsubsim algorithm used in \cite{Kulakova_2016} for calibration of Lennard-Jones potential of helium [Section \ref{sec:speedup} \& \ref{sec:resLJ}]. As noted in \cite{Kulakova_2016}, High Performance Computing (HPC) and parallelization are essential in the context of MD due to the long simulation time and chaotic and unstable behavior of more realistic and challenging force-field formalisms. This need is addressed thanks to our recently developed Python eco-system called ABCpy   \citep{Dutta_2017_PASC}, which implements and parallelizes most of the existing ABC algorithms and adapts them optimally to HPC infrastructure. 
Further, to mitigate the imbalance in simulation times of different MD simulation models for different values of force-fields parameters, we use a new dynamic allocation scheme for MPI, developed in \cite{dutta2017abcpyhpc}. 

Having found that the APMCABC is more efficient than the ABCsubsim algorithm to calibrate force-field of helium, where no Gaussianity assumption is made, we pursue using it to calibrate the force-field of a more complicated system  such as water.
In this paper we only address the issue of calibrating the parameters given a force-field formalism, under the assumption of no measurement uncertainty. 
For this purpose we consider a force-field formalism of water, specifically the rigid non-polarizable TIP4P force-field developed by \cite{Jorgensen1983}, with all bonds and angles constrained using the LINCS algorithm mentioned in Section~\ref{sec:forcefields}.
Water is universal in science and life. From protein folding   \citep{Brotzakis2016} to ion mobility \citep{Ohtaki1993} and from enzymatic activity \citep{Abel2008} to anti-freeze activity \citep{Brotzakis2018}, local and/or global water structure and dynamics regulates biological and physicochemical processes. Hence, force-field parametrization by targeting structural or dynamical properties of water is of great impact in biology and science. 
We illustrate the performance of the inferential scheme for simulated dataset, generated using GROMACS, and for experimental datasets, obtained under ambient conditions using Neutron diffraction    \citep{soper2000radial} and X-ray diffraction  \citep{skinner2013benchmark}. 

In Section~\ref{sec:forcefields}, we explain the Lennard-Jones potential for helium and TIP4P force-field formalism of water, describe the models used to forward-simulate dataset using LAMMPS and GROMACS 
correspondingly. A short introduction to Bayesian inference and approximate Bayesian computation used for inferring the force-field parameters is given in Section~\ref{sec:if}. We infer the posterior distribution of the parameters and their Bayes estimates given a simulated or experimentally obtained dataset, in Section~\ref{sec:result}. As the proposed Bayesian inferential scheme provides us with a posterior distribution of the parameters, we can quantify the model prediction uncertainty, by simulating the dataset from the MD simulation model using values of the parameters randomly drawn from the inferred posterior distribution. Finally, in Section~\ref{sec:pred_uncertainty}, we illustrate the ability to quantify prediction uncertainty by our inferential scheme for the experimental dataset of water, collected using Neutron and X-ray diffraction technology.

\section{Force-fields}
\label{sec:forcefields}
\subsection{Lennard-Jones potential of helium}
\label{sec:LJHelpmodel}
We first consider the calibration of the parameters for 6-12 Lennard-Jones potential of helium. The potential is given by
\begin{eqnarray}
&&V_{LJ}(\sigma_{LJ},\epsilon_{LJ})\\\nonumber
&&=\sum_i\sum_j4\epsilon_{LJ}\left(\left(\frac{\sigma_{LJ}}{r_{ij}}\right)^{12}-\left(\frac{\sigma_{LJ}}{r_{ij}}\right)^{6}\right)
\end{eqnarray}
where $\epsilon_{LJ}$ (zJ), $\sigma_{LJ}$ (nm)
\footnote{Note that in this paper, for the helium system simulated with LAMMPS we use zJ and nm units for the $\sigma_{LJ}$ and $\epsilon_{LJ}$, whereas for the TIP4P water simulated with GROMACS, we used the GROMACS conventional units of kJ mol$^{-1}$ and nm for the $\sigma_{TP}$ and $\epsilon_{TP}$ respectively. 
We select these units to be consistent with the papers we have chosen as our benchmark.}
and $r_{ij}$ (nm)
are correspondingly the depth of the potential well, the finite distance at which the inter-particle potential is zero and the distance between the $i$ and $j$ particles. Using data we want to calibrate parameters $\epsilon_{LJ}$ and $\sigma_{LJ}$.
Thus, for the purpose of the present study, 
the LJ model of potential,
which we denote by $\Model_{LJ}$, is only parametrized in terms of the two non-bonded force-field parameters $\sigma_{LJ}$ and $\epsilon_{LJ}$, which we jointly denote with
\[ \parametertheo= (\sigma_{LJ}, \epsilon_{LJ}). \]

If the initial values for coordinates of helium atoms in an enclosed space are known, we can forward simulate 
the coordinates of the atoms over time using model $\Model_{LJ}$ for given values of these parameters $\parametertheo = \parametertheo^*$:
\begin{eqnarray}
\label{eq:simulator_LJ}
\Model_{LJ} [\parametertheo = \parametertheo^*] \rightarrow \left\lbrace 
\left(\mbox{$\mathbf{R}$}(t) \right): \ t=0, \ldots, \obsendtime \right\rbrace
\end{eqnarray}
where $\mbox{$\mathbf{R}$}(t)$ are 
the position of the molecules at time $t$ in an
enclosed space. 
Following \cite{Kulakova_2016, shinoda2004rapid}, the forward simulation 
in the NPT-ensemble of $300 K$ temperature and atmospheric pressure,
is performed using LAMMPS \citep{plimpton1995fast} 
in a simulation box of $27.3\  \times 27.3\ \times 27.3\ nm$ with 1000 helium atoms, for $\obsstarttime$ = 0, $\obsendtime = 5\ ns$ and timesteps of $2\ fs$, after the system has been equilibrated for $2\ ns$. The damping frequency for temperature and pressure was set at $2\ ps$ and $20\ ps$ respectively.
Furthermore, the Lennard-Jones interactions cutoff and the mass of helium atom was fixed at $.639\ nm$ and $6.64 e^{-6}\  attogram$.

\subsection{TIP4P force-field of water}
\label{sec:tip4pmodel}
TIP4P force-field \citep{Jorgensen1983} of water is a four interaction site water force-field with an extra charge placed on a dummy atom, besides the charges existing on the oxygen and hydrogens. It has been parametrized to reproduce  the enthalpy of vaporization  \citep{vega2011simulating}. The waters formalism is shown in 
Eq. \ref{eq:forcefield3}

\begin{eqnarray}
\label{eq:forcefield3}
U_{noncov}&=&\sum\limits_{i}^{}\sum\limits_{j}^{}{  4\epsilon_{TP}\left[ \left( \frac{\sigma_{TP}}{r_{ij}} \right)^{12} -\left(\frac{\sigma_{TP}}{r_{ij}}\right)^6\right]} \\\nonumber
&+& \sum\limits_{i}^{}\sum\limits_{j}^{}\sum\limits_{\alpha}^{}\sum\limits_{\beta}^{}{\frac{q_{i\alpha}q_{j\beta}}{r_{ij}}} 
\end{eqnarray}
where the potential energy function $U_{noncov}$ (kJ mol$^{-1}$) is the sum of 6-12 Lennard-Jones 
and electrostatic interactions. 
The $r_{ij}$ (nm) term corresponds to the intermolecular distance between  $i$-th and $j$-th water molecules.
The Pauli repulsion and the van der Waals attraction is parametrized with terms $\propto$ $r^{-12}$ and $\propto$ $r^{-6}$ correspondingly. 
The Lennard-Jones interaction term includes $\sigma_{TP}$ (nm) and $\epsilon_{TP}$ (kJ mol$^{-1}$) 
parameters, respectively, the distance at which the inter-particle potential is zero and the value of the minimum energy.
Further we keep the bond length and angles fixed in TIP4P rigid water model.
Finally, the electrostatic  term in Eq.~\ref{eq:forcefield3} involves the charges of atoms, where $i$ and $j$ correspond to different water molecules and $\alpha$ and $\beta$ are the indices of the partial charges q (e) of each molecule.

For the purpose of the present study, 
the TIP4P model of force-field,
which we denote by $\Model_{TP}$, is parametrized in terms of the two non-bonded force-field parameters $\sigma_{TP}$ and $\epsilon_{TP}$, which we jointly denote with
\[ \parametertheo= (\sigma_{TP}, \epsilon_{TP}), \]
representing the repulsion and attraction of the Van der Waals forces. 
Although the proposed model contains additional parameters, (e.g., the charges of Eq.~\ref{eq:forcefield3}), we chose not to infer them and assume they are constant. However, we 
stress that the ABC method is not bound by the number of parameters to infer, and our choice is based on the  illustrative purpose of this paper.
If the initial values for coordinates of water molecules in an enclosed space are known, we can forward simulate 
the coordinates of the molecules over time using model $\Model_{TP}$ for given values of these parameters $\parametertheo = \parametertheo^*$:
\begin{eqnarray}
\label{eq:simulator_tpi4p}
\Model_{TP} [\parametertheo = \parametertheo^*] \rightarrow \left\lbrace 
\left(\mbox{$\mathbf{R}$}(t) \right): \ t=0, \ldots, \obsendtime \right\rbrace. 
\end{eqnarray}
where $\mbox{$\mathbf{R}$}(t)$ are 
the position of the molecules at time $t$ in an
enclosed space. 
The forward simulation is performed using MD (TIP4P implementation in GROMACS) for $\obsstarttime=0$, $\obsendtime=100$ $ps$ 
and timesteps of 2 $fs$. 
We first construct a simulation box of $2.5\ \times 2.5\ \times 2.5\ nm$ and of 515 water molecules. 
Then, after compiling the TIP4P force-field with $\parametertheo^*$, we perform an energy minimization, followed by an NPT simulation. 
In the energy minimization step, we 
use the steepest descend algorithm for energy minimization  \citep{jaidhanenergy}. 
We further use 
the LINCS algorithm \citep{hess1997lincs} for the bond and angle constraints and the Particle Mesh Ewald (PME) \citep{darden1993particle} with a cut-off of 1 $nm$ to treat the electrostatics. The cut-off for the Van der Waals interactions 
is 1 $nm$. In the NPT simulation we set the temperature and the pressure at 298 K and 1 atm correspondingly, using the stochastic velocity rescaling thermostat \citep{Bussi2006} and the Parrinello-Rahman barostat  \citep{Parrinello1981} respectively. The neighbor list is updated every 10 timesteps. The electrostatic and Van der Walls parameters and the LINCS algorithm for the bond and angle constraints are treated in the same way as in the energy minimization. 

\section{Inference framework}
\label{sec:if}
We use the term observed data, denoted by $\dataObs$, to refer to a dataset generated by some real-world process 
(e.g., experimental studies using X-ray and Neutron diffraction, 
that allow us to measure different properties of water molecules), 
and our goal is to learn values of the force-field parameters characterizing this process. 
Assuming that the first observation of the process occurs at time $\obsstarttime$ and the last  at time $\obsendtime$, the observed dataset 
is $\dataObs \equiv \lbrace \mbox{$\mathbf{R}$}(t): \obsstarttime, \ldots, \obsendtime  \rbrace$. 
From experimental studies it is not possible to track the time dependent position of water molecules, but we can learn their properties, e.g. different radial distribution functions, using different diffraction techniques. 

Here we address the question of calibrating force-fields for any available structural and dynamical properties, such as the radial distribution function and self-diffusion coefficient. To this aim, we develop  an approximate Bayesian inference scheme that allows us to quantify the uncertainty in the inferred model parameters, uncertainty which is inherent to the inferential process given the chaotic 
nature of the models described in Equations~\ref{eq:simulator_LJ}\&\ref{eq:simulator_tpi4p}. We only calibrate
force-fields targeting structural and dynamical properties, but we stress that this methodology can be used for any property which is available experimentally.

\subsection{Bayesian inference}
\label{sec:BI}
We can quantify the  uncertainty of the parameter $\parametertheo$ by 
its posterior distribution $p(\parametertheo|\data)$ given the observed dataset $\data = \dataObs$. The posterior distribution is obtained by Bayes' theorem as,
\begin{eqnarray}
p(\parametertheo|\dataObs) = \frac{\prior(\parametertheo)
 p(\dataObs|\parametertheo)}{m(\dataObs)},
\end{eqnarray}
where $\prior(\parametertheo)$, $p(\dataObs|\parametertheo)$ and $m(\dataObs) = \int\prior(\parametertheo)p(\dataObs|\parametertheo)d\parametertheo$ are, correspondingly, the prior distribution on the parameter $\parametertheo$, the likelihood function, and the marginal likelihood. The prior distribution $\prior(\parametertheo)$ enables to 
incorporate, in the inferential process, prior knowledge on the parameter values.
If the likelihood function could be evaluated, at least up to a normalizing constant, then the posterior distribution could be approximated by drawing a representative sample of parameter values from it using (Markov chain) Monte Carlo sampling schemes \citep{Robert2005}. Unfortunately,  the likelihood function induced by the TIP4P water and helium model is analytically intractable
because we do not assume Gaussianity in the model. 
In this setting, approximate Bayesian computation (ABC)  \citep{lintusaari2017fundamentals} offers a way to sample from the approximate posterior distribution and opens up the possibility of sound statistical inference on the parameter $\parametertheo$. In this paper we only focus on parameter estimation/calibration and
uncertainty quantification but we stress that ABC easily allows to also perform parameter hypothesis testing  and model selection.

\subsection{Approximate Bayesian computation (ABC)}
\label{sec:Inference_method}
Models that are easy to forward simulate, given values of the parameters, are called simulator-based models in the ABC literature and are used in a wide range of scientific disciplines to describe and understand different aspects of nature ranging from dynamics of sub-atomic particles  \citep{Martinez_2016} to evolution of human societies \citep{Turchin_2013} and formation of universes \citep{Schaye_2015}.
In the fundamental rejection ABC sampling scheme, we simulate a synthetic dataset $\datasim$ from the simulator-based model $\Model(\parametertheo)$ for a fixed parameter value, $\parametertheo$, and measure the closeness between $\datasim$ and $\dataObs$ using a pre-defined discrepancy measure $\distance(\datasim,\dataObs)$. Based on this discrepancy, ABC accepts the parameter value $\parametertheo$ when $\distance(\datasim,\dataObs)$ is less than a pre-specified threshold value $\threshold$. 

The intractable likelihood 
$p(\dataObs|\parametertheo)$
is approximated by 
$p_{\distance,\delta}(\dataObs|\parametertheo)$
for some choice of distance, $d$, and threshold,  $\threshold>0$, where 
\begin{eqnarray}
\label{eq:approx_lik}
p_{\distance,\threshold}(\dataObs|\parametertheo) \propto P(\distance(\datasim,\dataObs)<\threshold)
\end{eqnarray}
and, as a consequence, the sampled parameters follow the posterior distribution of $\parametertheo$ conditional on $\distance(\datasim,\dataObs)<\threshold$:
\begin{align*}
p_{\distance,\threshold}(\parametertheo|\dataObs) \propto P(\distance(\datasim,\dataObs)<\threshold)\prior(\parametertheo).
\end{align*}
For a better approximation of the likelihood function, computationally efficient sequential ABC algorithms \citep{Marin_2012} 
decrease the value of the threshold $\threshold$ adaptively while exploring the parameter space. 

In this manuscript, we consider two sequential ABC algorithms: ABCsubsim  \citep{Chiachio_2014} and APMCABC \citep{lenormand2013adaptive}. At the first step of these two algorithms, $N_{\mbox{sample}}$-many parameter values are randomly drawn from the prior distribution and the value of $\threshold$ is decreased adaptively depending on the pseudo data simulated from the model using those randomly sampled parameter values. In the next step, they produce $N_{\mbox{sample}}$-many parameter values approximately distributed from the distribution $p_{\distance,\threshold}(\parametertheo|\dataObs)$, for the adapted $\threshold$ value from last step and again decrease the $\threshold$ depending on the new samples. This procedure is continued $N_{\mbox{step}}$ many times or until some stopping criterion is reached. We note that the adapted $\threshold$ values at each step, is strictly decreasing and converges to zero, in turn improving the approximation to the posterior distribution. Here we denote the value of $\threshold$ at the final step as $\threshold_{\mbox{final}}$.

\subsection{Discrepancy, prior distribution and perturbation kernel}
\label{sec:abcdetial}
The discrepancy measure between $\datasim$ and $\dataObs$ is often defined through a distance between 
 summary statistics computed from $\datasim$ and $\dataObs$. 
 The choice of these summary statistics is a crucial aspect for a good ABC approximation.
 The summary statistics are usually chosen to minimize the loss of information on $\parametertheo$ contained in the data and picking low-dimensional summaries to avoid curse of dimensionality  \citep{fearnhead2012constructing}. Here we use intuitive discrepancy measures between interpretable 
and domain-driven summary statistics. Subjectivity of these decisions can be removed through automatic summary selection for ABC, described in \cite{fearnhead2012constructing, pudlo2015reliable, jiang2015learning, gutmann2017likelihood}, where an informative linear or non-linear combination of the summaries is chosen. 
Below we provide the summary statistics extracted from the dataset, the discrepancy measure,
the prior distribution of the parameters, and the perturbation kernel used to explore the parameter space in the APMCABC algorithm for the calibration of LJ potential and TIP4P force-field of helium and water,  correspondingly. The discrepancy measure used for LJ force-field of helium, was previously used  in \cite{Kulakova_2016}.

\subsubsection{LJ potential of helium}

\paragraph*{Summary statistics:} Given a dataset $\data \equiv \left\lbrace \mbox{$\mathbf{R}$}(t) : t = \obsstarttime , \ldots , \obsendtime \right\rbrace$ for LJ potential of helium simulated using LAMMPS, we compute the following summary statistics,
$$\summaryfunc_{LJ}: \data \rightarrow (\mathbf{f}_B \equiv f_B(t) : t = \obsstarttime , \ldots , \obsendtime)$$
where $f_B(t) = \langle exp\{-H(t)/(k_BT)\}\rangle$, $k_B$ is the Boltzmann constant, $T$ is the temperature of the system, $H(t)$ is the enthalpy contribution of a helium atom in the system at time $t$ and $\langle \ \rangle$ denotes the ensemble average over all the atoms in the system at a time instance $t$.

\paragraph*{Discrepancy measure:} The discrepancy measure between two datasets $\data^{(1)}$ and $\data^{(2)}$ is constructed by considering the Kullback-Leibler divergence between the probability distribution functions $\chi^{(1)}$ and $\chi^{(2)}$, of $\mathbf{f}^{(1)}_B$ and $\mathbf{f}^{(2)}_B$ over time instances extracted from $\data^{(1)}$ and $\data^{(2)}$. 
\begin{eqnarray*}
\label{eq:discrep_measure}
\distdata_{LJ}(\data^{(1)}, \data^{(2)})
&:=&\distdata_{LJ}\left(\summaryfunc_{LJ}(\data^{(1)}), \summaryfunc_{LJ}(\data^{(2)})\right)\\
&:=&\distdata_{LJ}\left(\mathbf{f}^{(1)}_B,\mathbf{f}^{(2)}_B\right)\\
&=& \int \chi^{(1)}(z) \log{\frac{\chi^{(1)}(z)}{\chi^{(2)}(z)}}dz.
\end{eqnarray*}
\paragraph*{Prior distributions:} 
For $\sigma_{LJ}$ and $\epsilon_{LJ}$ we use independent continuous uniform prior distributions on the range  $[0.1(nm), 0.8(nm)]$ and $[0.01(zJ), 1.0(zJ)]$, respectively. 

\paragraph*{Perturbation kernel:} To explore the parameter space of $\parametertheo = (\sigma_{LJ}, \epsilon_{LJ}) \in 
[0.1(nm), 0.8(nm)]\times[0.01(zJ), 1.0(zJ)$],
we consider a truncated two-dimensional multivariate Gaussian distribution on the above space as the perturbation kernel. APMCABC inference scheme centers the perturbation kernel at the parameter value it is perturbing and updates the variance-covariance matrix of the perturbation kernel based on the parameter values sampled from the previous step.

\subsubsection{TIP4P force-filed of Water}

\begin{figure}[h!]
  \centering
  \subfloat[radial distribution function
of $O-H$]{\includegraphics[width=0.45\textwidth]{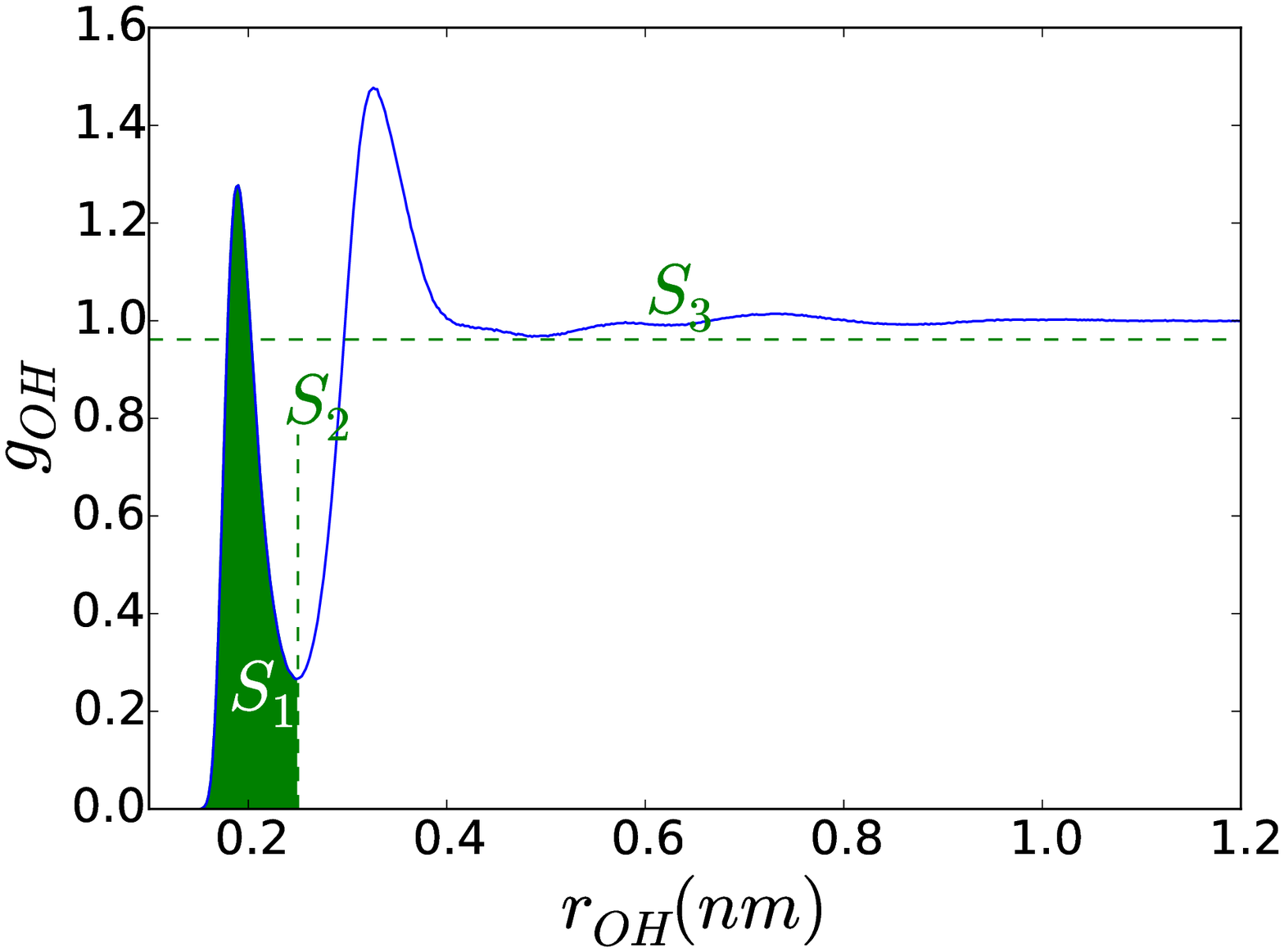}\label{fig:rdfoh}}
  \hfill
  \subfloat[radial distribution function
of $O-O$]{\includegraphics[width=0.45\textwidth]{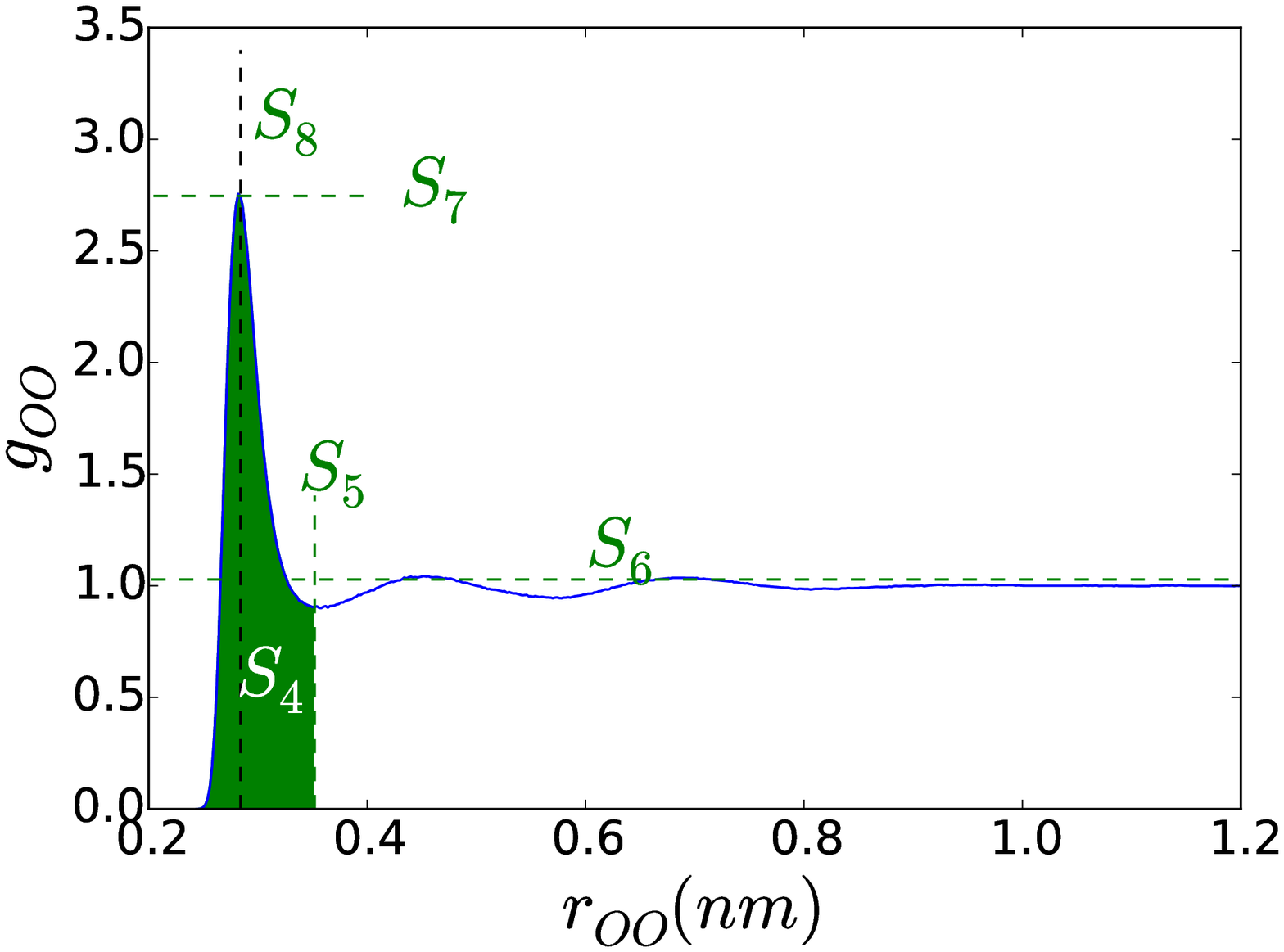}\label{fig:rdfoo}}
   \hfill
     \subfloat[mean square displacement]{\includegraphics[width=0.45\textwidth]{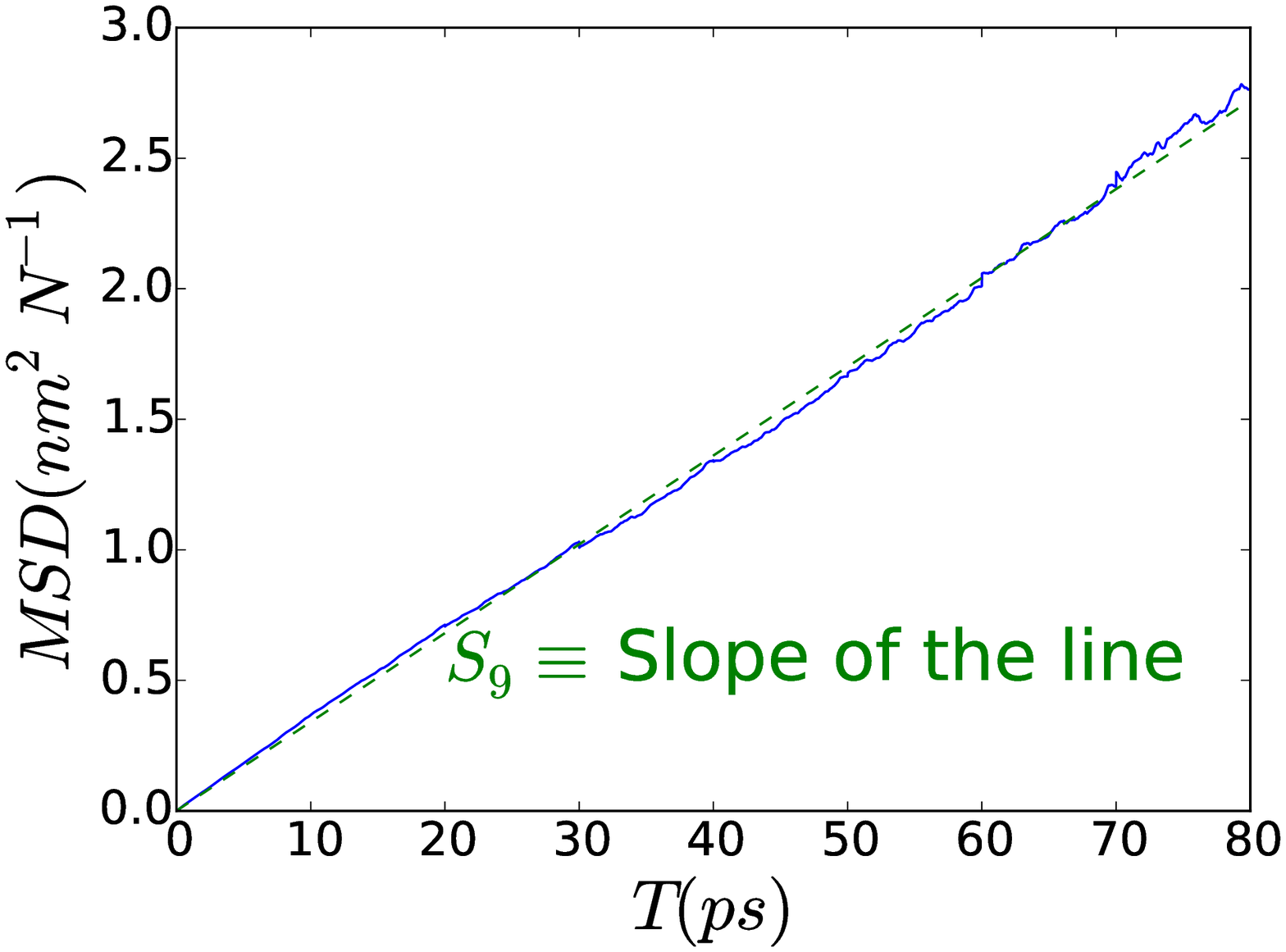}\label{fig:msd}}
  \caption{\textbf{Summary statistics (Water):}  $\left( S_1, S_2, S_3, S_4, S_5, S_6, S_7, S_8, S_9 \right)$ 
  are computed from the radial distribution functions
for $O-H$ \textbf{(a)}, $O-O$ \textbf{(b)} atoms and the mean square displacement \textbf{(c)}, generated from the simulated coordinates of the dynamical system.}
  \label{fig:summ_stat}
\end{figure}

\paragraph*{Summary statistics:} 
Given a dataset $\data \equiv \left\lbrace \mbox{$\mathbf{R}$}(t) : t = \obsstarttime , \ldots , \obsendtime \right\rbrace$ for TIP4P model of water simulated using GROMACS, we compute an array of summary statistics,
$$\summaryfunc_{TP}: \data \rightarrow (S_1,S_2,S_3,S_4,S_5,S_6,S_7, S_8,S_9)$$
defined as follows:
\begin{itemize}
\item $S_1$: Estimate of the number of hydrogen bonds per water molecule - The area under the curve $r_{OH}$ vs $g_{OH}$ until the first minimum;
\item $S_2$: Estimate of the donor acceptor hydrogen bond distance - Value of $r_{OH}$ (nm) at the first minimum of the radial distribution function $g_{OH}$;
\item $S_3$: Mean of $g_{OH}$;
\item $S_4$: Estimate of number of water molecules in the first hydration shell - The area under the curve $r_{OO}$ vs $g_{OO}$ until the first minimum;
\item $S_5$: Estimate of the maximum distance of the first hydration shell - Value of $r_{OO}$ (nm) at the first
minimum of the radial distribution function $g_{OO}$;
\item $S_6$: Mean of $g_{OO}$;
\item $S_7$: The height of $g_{OO}$ at the first maximum of $g_{OO}$;
\item $S_8$: Value of $r_{OO}$ (nm) at the first
maximum of the radial distribution function $g_{OO}$;
\item $S_9$: Slope of the line, fitted to (M)ean (S)quare (D)isplacement (MSD), which is an estimate of 6 $\times$ self-diffusion coefficient. The MSD is in units of nm$^2$N$^{-1}$.
\end{itemize}
To compute the above summary statistics, we first compute the radial distribution functions
for the $O-H$ and $O-O$ atoms and the mean square displacement (MSD) from the simulated
coordinates of the dynamical system. We then compute the above mentioned summary statistics from the radial distribution functions and the MSD as explained in  Figure~\ref{fig:summ_stat}.
The intuition behind choosing the above mentioned quantities is that they are broadly used characteristic quantities of the structure and dynamics of liquids \citep{Allen1989,Frenkel2001}.

\paragraph*{Discrepancy measure:} The discrepancy measure between two datasets $\data^{(1)}$ and $\data^{(2)}$ is constructed by considering the distance functions between the summary statistics extracted from them:
\begin{eqnarray*}
\label{eq:discrep_measure}
\distdata_{TP}(\data^{(1)}, \data^{(2)})
&:=&\distdata_{TP}\left(\summaryfunc_{TP}(\data^{(1)}), \summaryfunc_{TP}(\data^{(2)})\right)\\
&=& \frac{1}{9}\sum_{i=1}^9|S_i^{(1)}-S_i^{(2)}|
\end{eqnarray*}

\paragraph*{Prior distributions:} 
We use independent continuous uniform prior distributions on the range 
$[0.281(nm), 0.53(nm)]$ and $[0.2(kJ mol^{-1}), 0.9(kJ mol^{-1})]$ correspondingly for $\sigma_{TP}$ and $\epsilon_{TP}$. 
Outside this parameter range the TIP4P model of water in GROMACS becomes extremely chaotic and 
simulated data set can not be obtained  in a reasonable time span.

\paragraph*{Perturbation kernel:} The perturbation kernel used to explore the parameter space of $\parametertheo = (\sigma_{TP}, \epsilon_{TP}) \in 
[0.281(nm), 0.53(nm)]\times[0.2(kJ mol^{-1}), 0.9(kJ mol^{-1})$], 
is a truncated two-dimensional multivariate Gaussian distribution. APMCABC inference scheme centers the perturbation kernel at the parameter value it is perturbing and updates the variance-covariance matrix of the perturbation kernel based on the parameter values sampled in the previous step.

\subsection{Time imbalance and dynamic allocation for MPI}
\label{sec:dynamicMPI}

\begin{figure}[htbp]
  \centering
    \subfloat[Lennard-Jones potential of helium]{\includegraphics[width=0.45\textwidth]{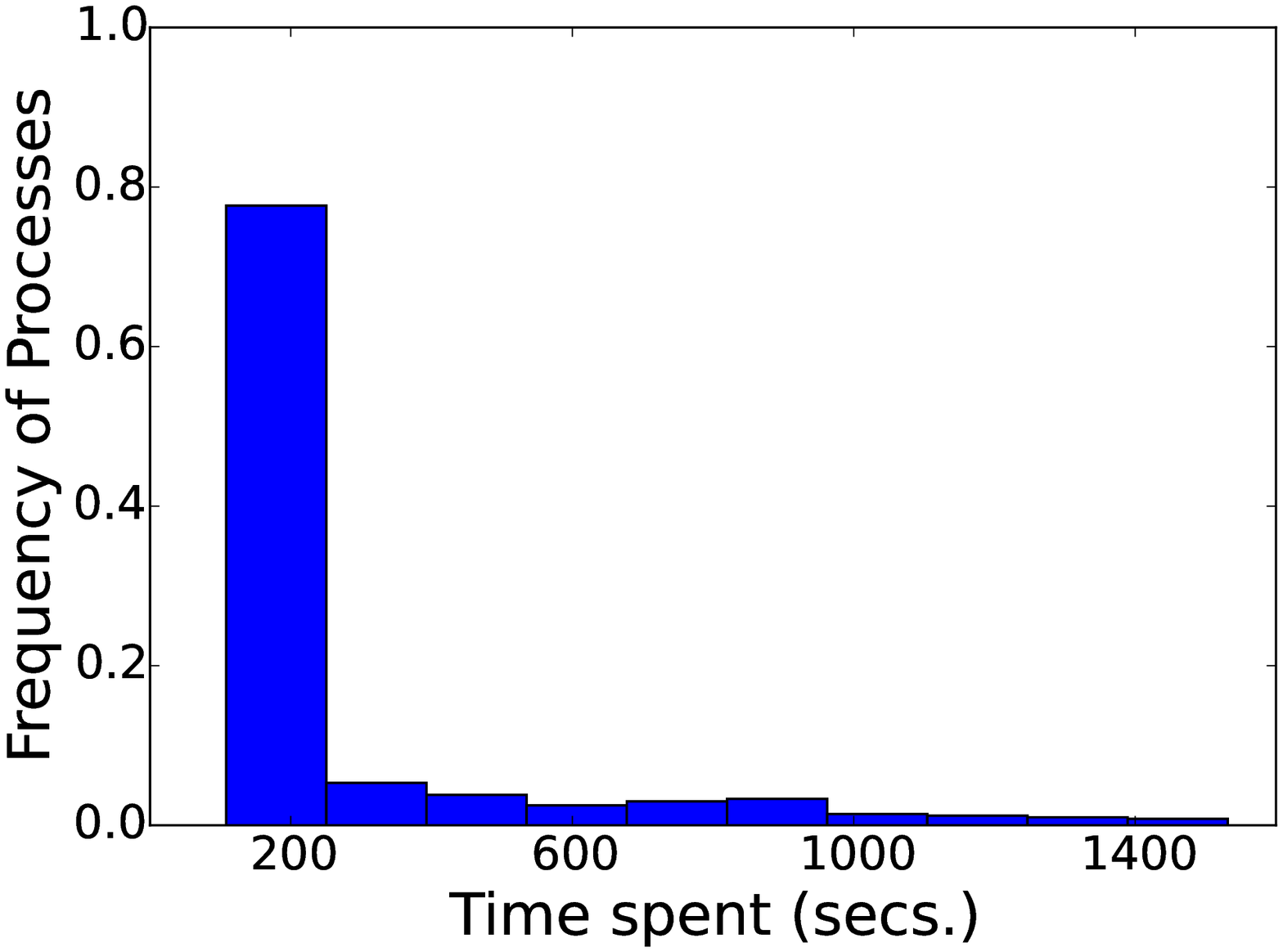}\label{fig:imbalance_helium}}
    \hfill
      \subfloat[TIP4P force-field of Water]{\includegraphics[width=0.45\textwidth]{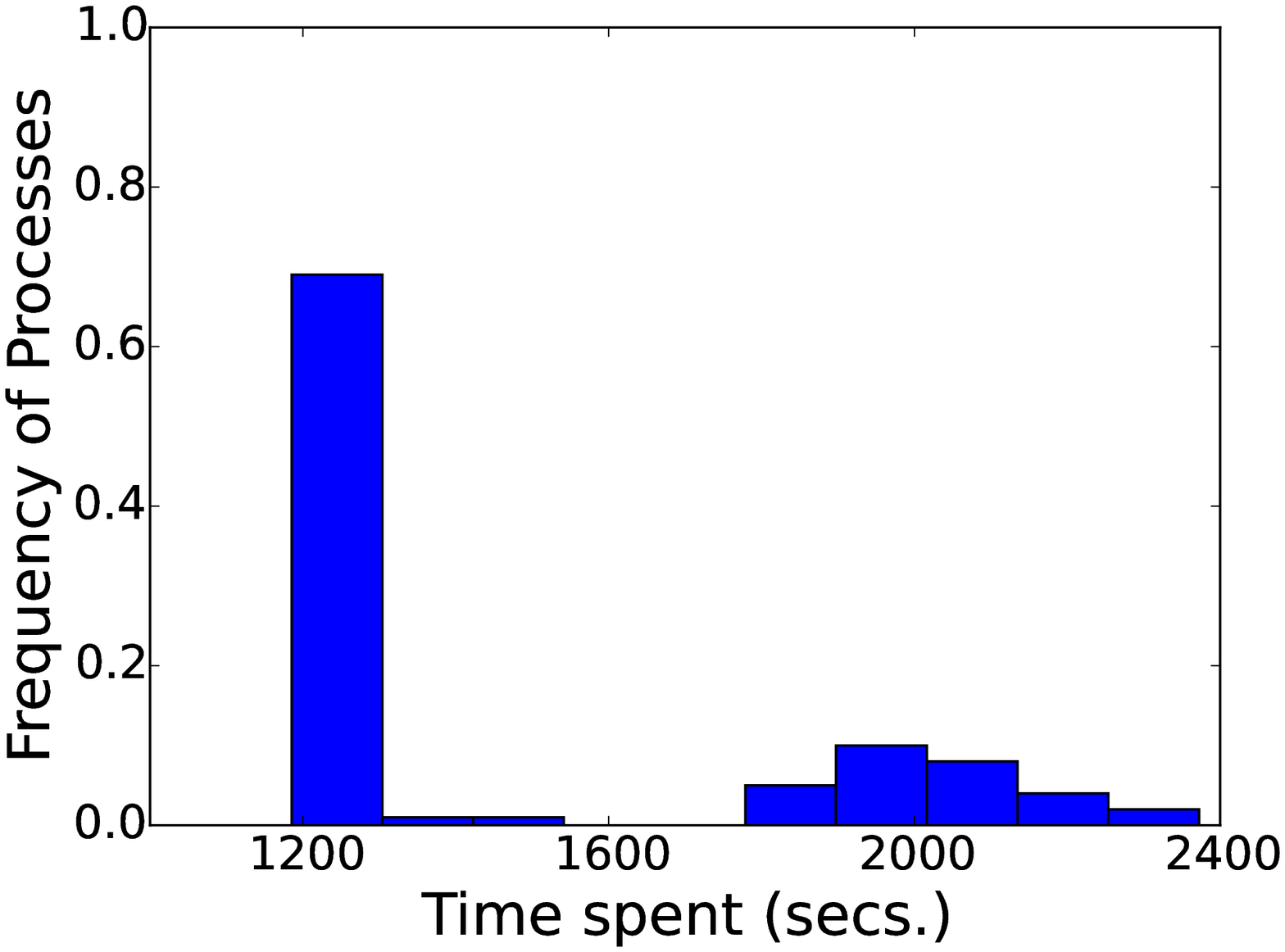}\label{fig:imbalance_water}}
  \caption{Imbalance in time spent to simulate a pseudo data for different values of $\parameter$.}
  \label{fig:imbalance}
\end{figure}

\begin{figure}[h]
  \centering
  \subfloat[MPI Backend]{\includegraphics[width=0.45\textwidth]{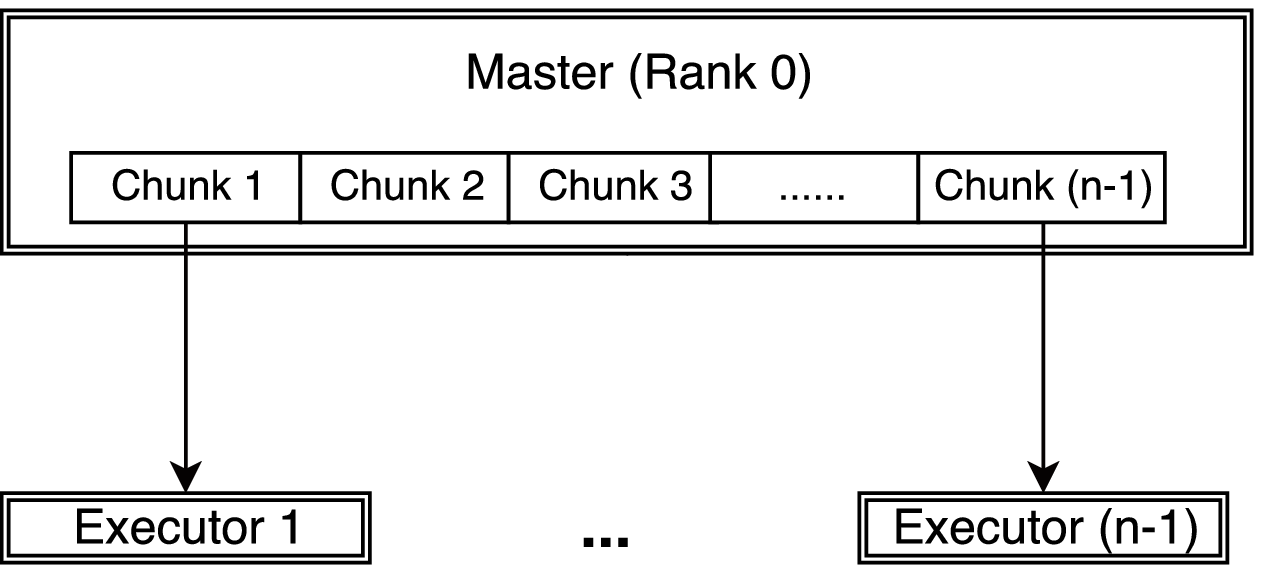}\label{fig:MPI_workflow_basic}}
  \hfill
  \subfloat[dynamic-MPI Backend]{\includegraphics[width=0.45\textwidth]{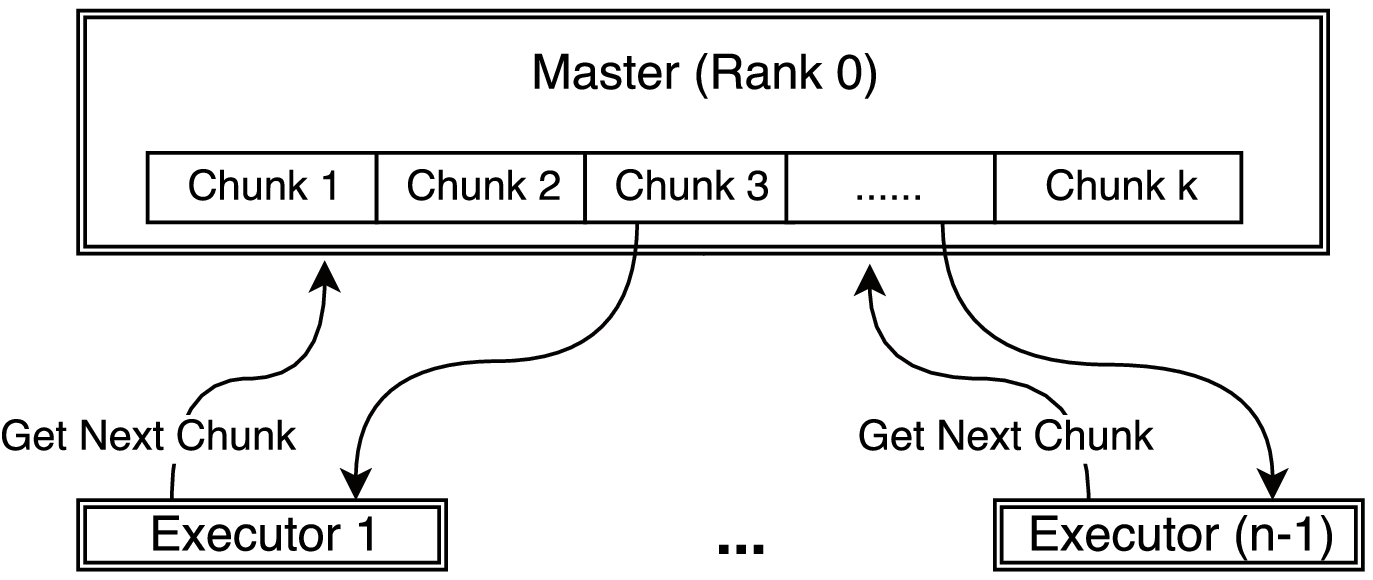}\label{fig:MPI_workflow_dynamic}}
   \hfill
  \caption{Comparison of work-flow between MPI and dynamic-MPI backend.}
  \label{fig:MPI_workflow_comparison}
\end{figure}

For inference, we use ABCpy Python package \citep{Dutta_2017_PASC}, which implements some of the most advanced ABC algorithms with an optimal exploitation of an HPC environment. 
The parallelization schemes
in ABCpy were primarily meant for inferring parameters from models which take almost
equal time to simulate dataset for any values of $\parametertheo$. Though, due to the chaotic nature of
the MD systems, we observe in Figure~\ref{fig:imbalance} that, for different values of $\parametertheo$, 
the simulation-time, for a  fixed $\obsendtime$ value, is quite variable 
(with 36-core of Piz Daint Cray architecture: Intel Broadwell with NVidia TESLA P100).
To solve this imbalance, we use a new dynamic allocation scheme for MPI, developed in \cite{dutta2017abcpyhpc}.

Here, we briefly explain the dynamic work allocation strategy for map-reduce in comparison to a straightforward allocation approach. In the straightforward approach, the allocation scheme initially distributes $m$ tasks to $n$ executors, sends the map function to each executor, which in turn applies the map function, one after the other, to its $m/n$ map tasks. This approach is visualized in Figure~\ref{fig:MPI_workflow_basic}, where a chunk represents the set of $m/n$ map tasks. For example, if we want to draw
$10,000$ samples from the posterior distribution and we have $n = 100$ cores available, at each step of APMCABC we create groups of 100 parameters and each group is assigned to one individual core.

On the other hand, the dynamic allocation scheme initially distributes $k < m$ tasks to the $k$ executors, sends the map function to each executor, which in turn applies it to the
single task available. In contrast to the straightforward allocation, the executor requests a new map task as soon as the previous one is terminated. This clearly results in a better work balance. The dynamic allocation strategy is an implementation of the famous greedy algorithm for job-shop scheduling, which can be shown to have an overall processing time (makespan) up to twofold better than the best makespan    \citep{Graham1966}. This approach is illustrated in Figure~\ref{fig:MPI_workflow_dynamic}.
The unbalancedness is not a problem that can be overcome easily by adding resources, rather speed-up and efficiency can drop drastically compared to the dynamic allocation strategy with increasing number of executors. For a detailed description and comparison, we direct readers to \cite{dutta2017abcpyhpc}.

\subsection{Efficient scale-up using adaptive population Monte Carlo ABC}
\label{sec:speedup}
\begin{figure}[h!]
  \centering
     \subfloat[Performance: speedup]{\includegraphics[width=0.45\textwidth]{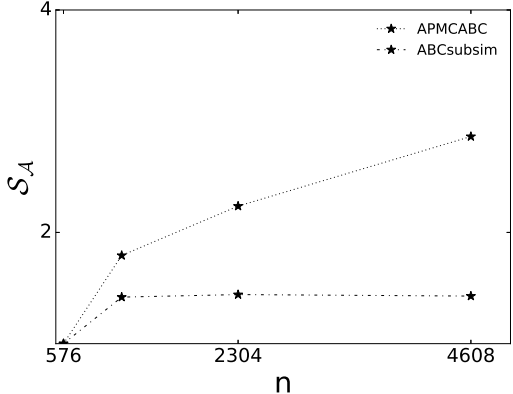}\label{fig:speedup}}
  \caption{\textbf{Performance comparison:}
Computational speedup $\mathcal{S}_{\mathcal{A}}(n)$ of APMCABC versus ABCsubsim algorithm for calibration of Lennard-Jones force-field of helium 
using dynamic-MPI backend with different number of cores, $n$.}
  \label{fig:comp_abcsubsim_apmcabc}
\end{figure}

Given the time imbalance and the high simulation cost of obtaining a pseudo dataset from $\Model_{TP}$, we need to choose an inference algorithm which converges fast with few number of pseudo data simulation and can thus 
better exploit the available computational resources. 
ABCsubsim algorithm was used for calibration of Lennard-Jones force-filed of helium in \cite{Kulakova_2016}. 
Here, we propose instead to use adaptive population Monte Carlo ABC (APMCABC) \citep{lenormand2013adaptive} algorithm. 

In Figure~\ref{fig:comp_abcsubsim_apmcabc}, we compare the performance of ABCsubsim with APMCABC algorithm, with reference to the speedups obtained while increasing the number of cores. An higher-order scale-up in performance can be noted for APMCABC algorithm. The discrepancy in speedup can be explained by the purely sequential nature of multiple Metropolis-Hastings chains used to update the samples at each steps of ABCsubsim, compared to purely parallel nature of APMCABC at each step, where the main job can be divided into $N_{\mbox{sample}}$-many parallel jobs. 

Hence, in this paper, we choose APMCABC  for the calibration of force-fields. Later in Section~\ref{sec:resLJ}, we also compare these two algorithms depending on their achieved threshold ($\threshold$) value and Bayes estimate after a fixed number of steps for the calibration of helium system, illustrating a faster convergence gained by APMCABC. 
Thanks to APMCABC, we can approximate the posterior distribution $p(\parametertheo|\dataObs)$ by drawing samples from it. 

\subsection{Parameter estimation}
\label{sec:parameter_estimation}

Our main goal is to estimate $\parametertheo$, given
$\dataObs$. In decision theory, the Bayes estimator minimizes the posterior expected loss, $E_{p(\parametertheo|\dataObs)}(\lossfunc(\parametertheo,\cdot)|\dataObs)$. Here we consider the following loss function,\\
\begin{align*}
\lossfunc(\parametertheo_1,\parametertheo_2) := \eucldist(\phi_1,\phi_2)
\end{align*}
were $\eucldist$ is the Euclidean distance. If we have $Z$ 
samples $(\parametertheo_{i})_{i=1}^{Z}$ from the posterior distribution $p(\parametertheo|\dataObs)$, the Bayes estimator can be 
approximated by
\begin{eqnarray}
\label{eq:Bayes_estimate}
\estparametertheo = \argmin_{\parametertheo} \frac{1}{Z}\sum_{i=1}^Z \lossfunc(\parametertheo_{i},\parametertheo)
\end{eqnarray}
which is also the estimated posterior mean $\estparametertheo = \frac{1}{Z}\sum_{i=1}^Z \parametertheo_{i}$.

\section{Result}
\label{sec:result}
We now illustrate how the Bayesian inference scheme introduced in Section~\ref{sec:if}, 
can be used to infer the posterior distribution and Bayes estimates of parameter $\parametertheo$ given an observed dataset $\dataObs$ both in simulated and experimental settings. 

\subsection{LJ potential of helium}
\label{sec:resLJ}
\subsubsection*{Simulated Data.}
\begin{figure}[htbp]
  \centering
\includegraphics[width=.45\textwidth]{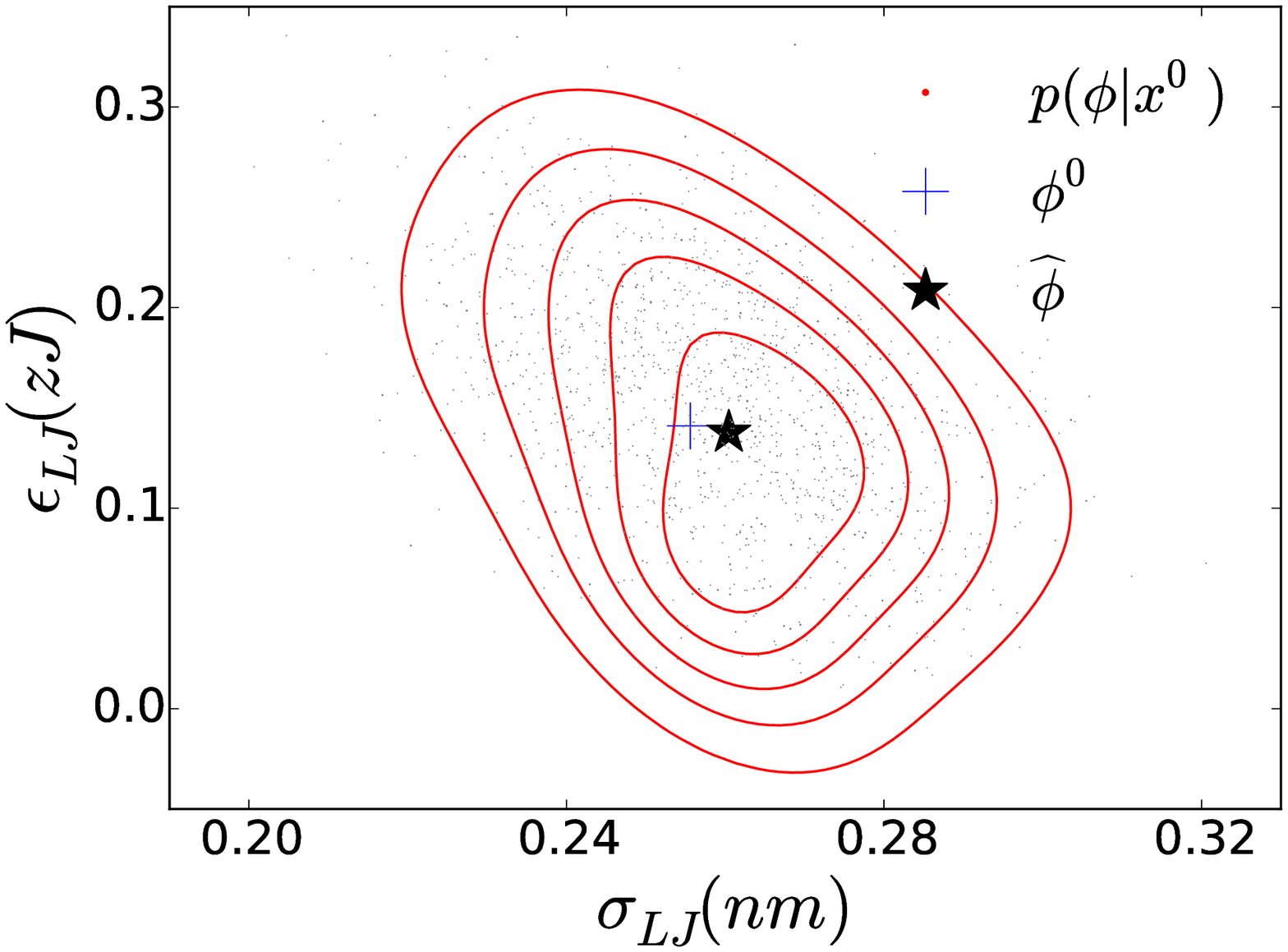}
  \caption{\textbf{Inference on Simulated Data (Helium):} Contour plot of posterior distribution $p(\parametertheo|\dataObs)$ (red), Bayes estimate $\widehat{\parametertheo}\equiv (\widehat{\sigma}_{LJ},\widehat{\epsilon}_{LJ}) = (0.26(nm),0.137(zJ))$ (black star) and  true value $\parametertheo^0\equiv (\sigma_{LJ}^0,\epsilon_{LJ}^0) = (0.255(nm),0.141(zJ))$ (blue cross) used to generate the simulated dataset. The posterior correlation between the parameters is $-0.38$. The posterior distribution is obtained using a Gaussian kernel density estimator with bandwidth $0.7$}
  \label{fig:helium_sim}
\end{figure}

The simulated data was generated from
$\Model_{LJ}$ using the software LAMMPS for a
system consisting of 1000 atoms, with $\obsendtime = 20\ ns$ and a timestep of $2\ fs$, under NPT conditions. 
To generate the simulated data that, in our proof of concept, will play the role of observed data, 
the parameter values were fixed at $\parametertheo^0 \equiv (\sigma_{LJ}^0,\epsilon_{LJ}^0) = (0.2556(nm), 0.141(zJ))$ as in \cite{Kulakova_2016}. We note that the value 
of $\obsendtime$ simulation model is $5\ ns$ compared to $ t_{end}=20\ ns$ for the simulated dataset. As explained in \cite{Kulakova_2016}, the simulation model cannot replicate exactly the target data due to the smaller sampling time and this causes uncertainty due to modeling error, in addition to computational and parametric uncertainty. 
For inference we use APMCABC algorithm, with the tuning parameters fixed at the default values recommended in the ABCpy package with the exception of the $N_{\mbox{sample}}$, $N_{\mbox{step}}$ and the acceptance rate cut-off, which are set to 5000, 6 and 0.03, respectively. 
In Figure~\ref{fig:helium_sim}, we illustrate the inferred 
posterior distribution and the Bayes estimate $(\widehat{\parametertheo})$ which is in close agreement with the true parameter values $(\parametertheo^0)$ used to simulate the dataset. 

ABCsubsim algorithm was used to infer the parameters of the above problem using the exact same setup (eg. simulated dataset, discrepancy measure, prior distribution etc.) as in \cite{Kulakova_2016}. In Table~\ref{table:compare_algo}, we compare the performance of APMCABC and  ABCsubsim, by comparing the finally achieved threshold value ($\threshold_{\mbox{final}}$) and the Euclidean distance of the Bayes estimate from the true value after running APMCABC for $N_{\mbox{step}}=6$, 
with the ones reported in \cite{Kulakova_2016} for ABCsubsim. We note that APMCABC achieves a much smaller $\threshold_{\mbox{final}}$ value of 0.0138 compared to the value 0.67, achieved by ABCsubsim after $N_{\mbox{step}}=6$. Further the Bayes estimate learned using the inferred posterior samples of APMCABC is in better agreement with the true parameter values used to simulate the dataset, as illustrated by a smaller value of $d_E(\widehat{\parametertheo},\parametertheo^0)$, the Euclidean distance of the Bayes estimate from the true parameter value used to simulate the dataset. This shows a faster convergence and superior inferential performance of APMCABC algorithm compared to ABCsubsim, in addition to the better speedups achieved in Figure~\ref{fig:comp_abcsubsim_apmcabc}. 

\begin{table}[h]
\begin{minipage}{\textwidth}
\caption{\textbf{Comparison of Inference (Helium):} Euclidean distance of the Bayes estimate from the true parameter value used to simulate the dataset $d_E(\widehat{\parametertheo},\parametertheo^0)$ and the final threshold value ($\threshold_{\mbox{final}}$) achieved by APMCABC and ABCsubsim algorithms for calibration of Lennard-Jones force-field of helium, after $N_{\mbox{step}}=6$.}
\label{table:compare_algo}
\begin{center}
\begin{tabular}{r|c|c|c}
Algorithm & $d_E(\widehat{\parametertheo},\parametertheo^0)$ & $N_{\mbox{step}}$ & $\threshold_{\mbox{final}}$\\\hline
APMCABC & 0.00744 & 6 & 0.0138\\
ABCsubsim & 0.03365 & 6 & 0.67\\\hline
\end{tabular}
\end{center}
\end{minipage} 
\end{table}

\subsection{TIP4P force-field of Water}
\label{sec:resTP}
\subsubsection*{Simulated Data.}
We first consider a simulated setting where the observed data has been generated, using GROMACS, from the TIP4P model $\mathcal{M}_{\mbox{TP}}$ of water reported in Equation~\ref{eq:simulator_tpi4p}. To generate the simulated observed data, we fix the non-bonded force-field parameter values at $\parametertheo^0 \equiv (\sigma_{TP}^0,\epsilon_{TP}^0) = (0.315(nm), 0.648(kJ mol^{-1}))$, which has been found to accurately reproduce the heat of
vaporization, critical density, temperature and liquid density at 298K   \citep{Jorgensen1983}. As the true parameter value $\parametertheo^0$ is known, we can assess the performance of the posterior distribution and the Bayes estimate, correspondingly, by their concentration and closeness to $\parametertheo^0$. 

\begin{figure}[htbp]
  \centering
\includegraphics[width=.45\textwidth]{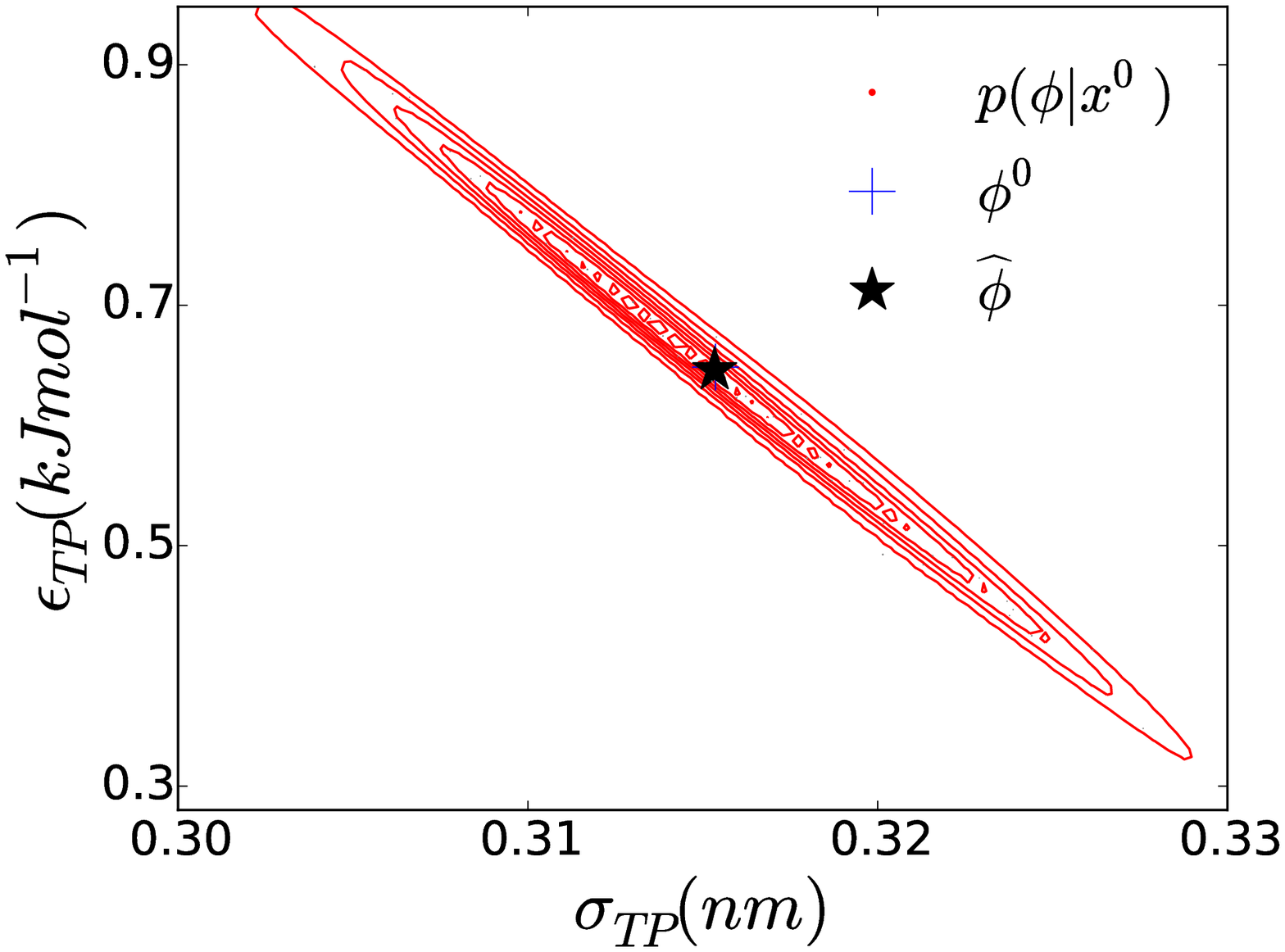}
  \caption{\textbf{Inference on Simulated Data (Water):} Contour plot of posterior distribution $p(\parametertheo|\dataObs)$ (red), Bayes estimate $\widehat{\parametertheo}\equiv (\hat{\sigma}_{TP},\hat{\epsilon}_{TP}) = (0.3153(nm),0.646(kJ mol^{-1}))$ (black star) and  true value $\parametertheo^0\equiv (\sigma_{TP}^0,\epsilon_{TP}^0) = (0.315(nm),0.648(kJ mol^{-1}))$ (blue cross) used to generate the simulated dataset.  
  A strong negative posterior correlation of $-0.994$ between the parameters is present. The posterior distribution is obtained using a Gaussian kernel density estimator with bandwidth $0.7$.}
  \label{fig:md_apmcabc_sim}
\end{figure}

For inference, all the tuning parameters of the APMCABC algorithm are fixed at the default values in the ABCpy package with the exception of $N_{\mbox{sample}}$, $N_{\mbox{step}}$ and the acceptance rate cut-off, which are set to 100, 10 and 0.03, correspondingly. 
In Figure~\ref{fig:md_apmcabc_sim}, the inferred posterior distribution, Bayes estimate ($\widehat{\parametertheo}$) and the true parameter value ($\parametertheo^0$) clearly show a very good performance of our inference scheme in estimating the underlying parametrization and quantifying the uncertainty in the inference. 

Here we would like to note that the summary statistics $(S_1,S_2,...,S_9)$
may have not converged for each of the parameter values at $t_{end}=100ps$, contributing to the Modeling uncertainty, as $t_{end}=100ps$ is part of our model specification. The concentrated nature of the posterior distribution in Figure 6, 
shows, however, a negligible presence of modeling uncertainty and we can implicitly conclude that the values of these statistics did converge for most of the considered parameter values in the range of prior distribution.
Additionally, to point the strength of having a posterior distribution for the parameters, we compute the posterior correlation between $\sigma_{TP}$ and $\epsilon_{TP}$, highlighting a strong negative correlation of $-0.994$.

\subsubsection*{Experimental Data}
We now illustrate the performance of the inference scheme for experimental dataset of water molecules under ambient conditions (temperature and pressure being fixed at 298K and 1 atm), assuming TIP4P force-field formalism of water. 
Though the exact coordinates over time of water molecules in an enclosed space can not be observed, the radial distribution functions of different molecular bonds and self-diffusion coefficient can be learned by experimental studies \citep{soper2000radial, skinner2013benchmark}. As our inference scheme only depends on the summary statistics extracted from the radial distribution functions of $O-O$, $O-H$ and self-diffusion coefficient, provided we have access to the experimentally obtained radial distribution functions and self-diffusion coefficient, we can 
compute the summary statistics and thus infer the non-bonded force-field parameters. 

\begin{figure}[htbp]
  \centering
    \subfloat[radial distribution function
of $O-H$]{\includegraphics[width=0.45\textwidth]{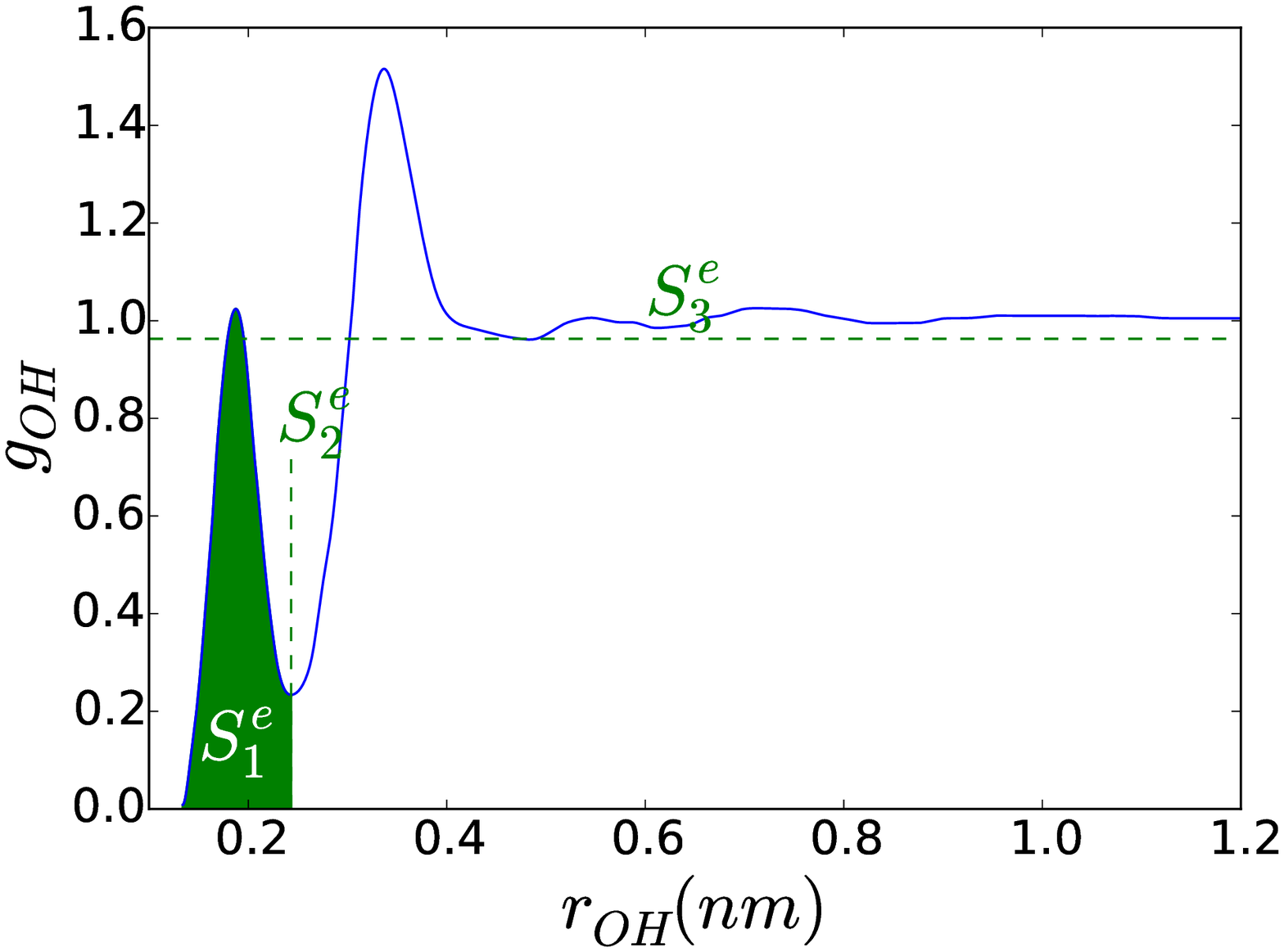}\label{fig:exprdfOH}}
    \hfill
      \subfloat[radial distribution function
of $O-O$]{\includegraphics[width=0.45\textwidth]{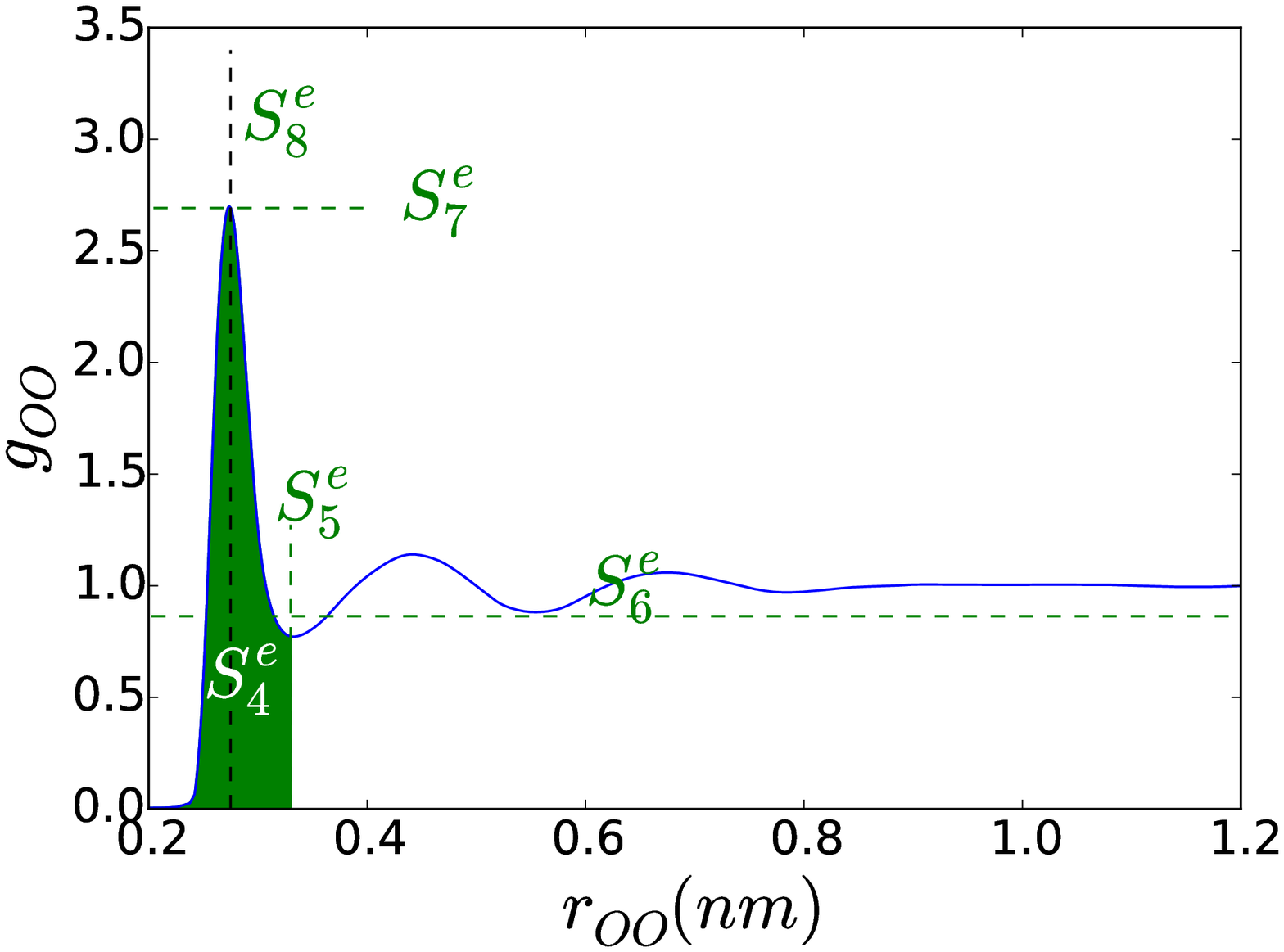}\label{fig:exprdfOO}}
    \hfill
      \subfloat[Inference on Neutron diffraction dataset]{\includegraphics[width=0.45\textwidth]{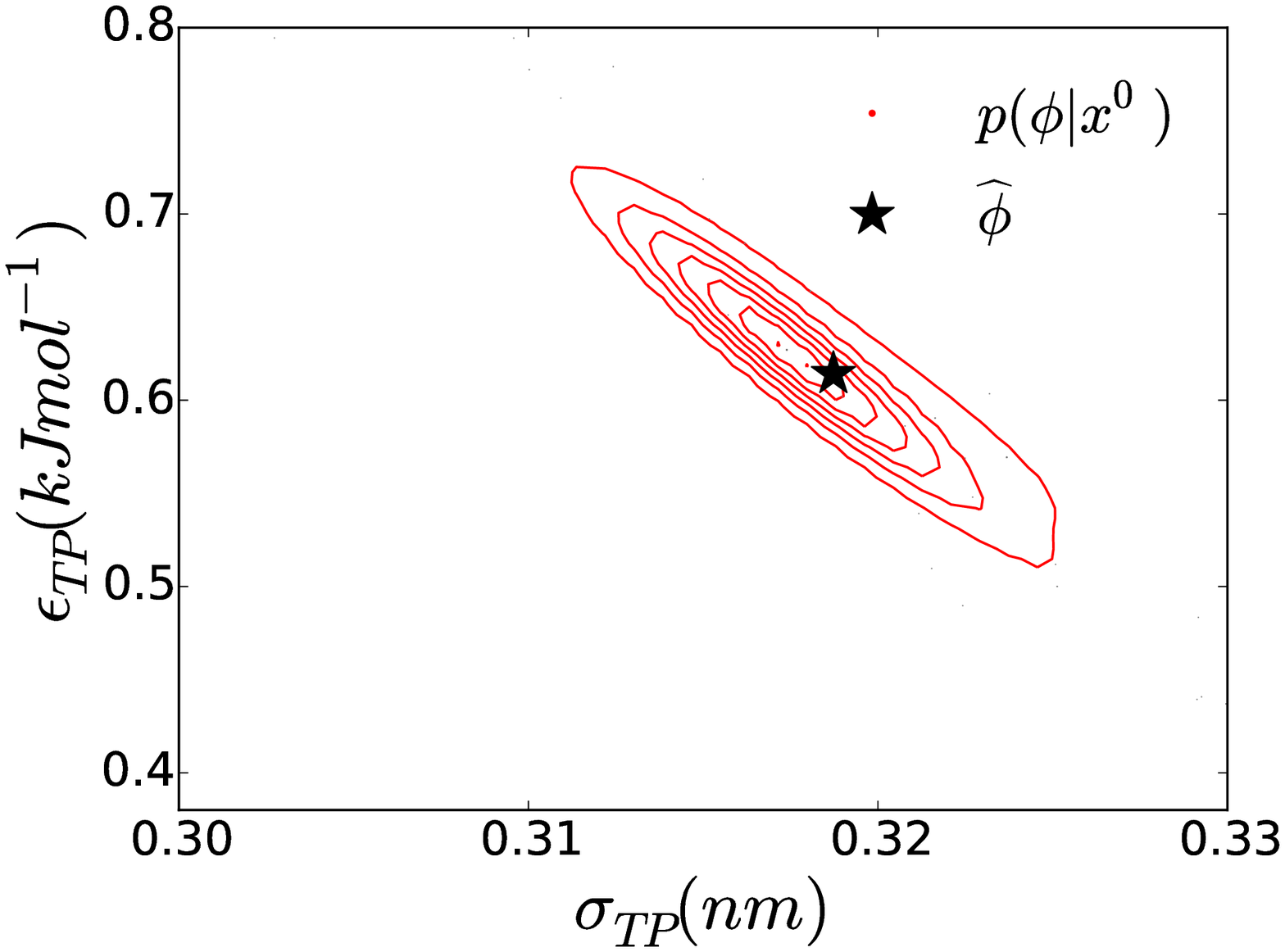}\label{fig:md_apmcabc_exp}}    
  \caption{\textbf{Neutron diffraction dataset (Water):} \textbf{(a)} \& \textbf{(b)} The radial distribution functions of $O-H$ and $O-O$ of water obtained by Neutron diffraction and reported in \cite{soper2000radial}. \textbf{(c)} Contour plot of posterior distribution $p(\parametertheo|\dataObs)$ (red), Bayes estimate $\widehat{\parametertheo}\equiv (\hat{\sigma}_{TP},\hat{\epsilon}_{TP}) = 
  (0.319(nm), 0.614(kJ mol^{-1}))$ (black star). A strong negative posterior correlation of $-0.95$ between the parameters is present. The posterior distribution is obtained using a Gaussian kernel density estimator with bandwidth $0.7$.}
  \label{fig:neutron_data}
\end{figure}

\paragraph*{Neutron Diffraction Dataset.}
First we consider the Neutron diffraction derived radial distribution function of water from \cite{soper2000radial}. The experimentally obtained radial distribution function of $O-O$ and $O-H$ are shown in Figure~\ref{fig:exprdfOH} \& \ref{fig:exprdfOO}, for further details we point readers to \cite{soper2000radial}. Additionally, we use the value of the self-diffusion coefficient, $1.3e^{-5} cm^2 s^{-1}$, reported in \cite{vega2011simulating}. The experimental values of the other summary statistics calculated are:
$S_1^{e} = 0.9628$, $S_2^{e}= 0.2432(nm)$, $S_3^{e}= 0.059$, $S_4^{e} = 0.8634$, $S_5^{e}= 0.33(nm)$, $S_6^{e}= 0.130$ and $S_7^{e}= 2.69$, $S_8^{e}= 0.275(nm)$. 

In Figure~\ref{fig:md_apmcabc_exp}, we plot the posterior distribution of $\parametertheo$ and the Bayes estimates $\widehat{\parametertheo}\equiv (\hat{\sigma}_{TP},\hat{\epsilon}_{TP}) = (0.319(nm), 0.614(kJ mol^{-1}))$ obtained using the proposed  inference scheme. The posterior correlation between $\sigma_{TP}$ and $\epsilon_{TP}$, has a negative correlation of $-0.95$  as in the case of simulated dataset. 

\begin{figure}[htbp]
  \centering
      \subfloat[radial distribution function
of $O-O$]{\includegraphics[width=0.45\textwidth]{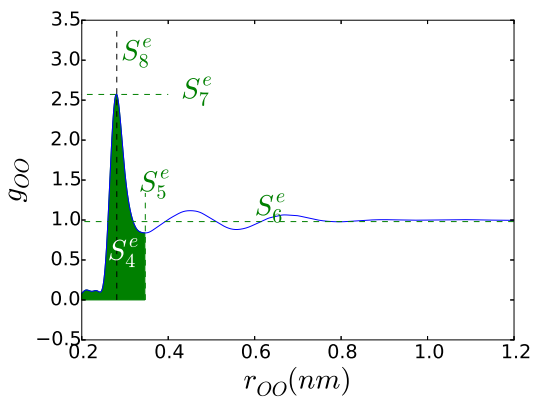}\label{fig:exprdfOO_2}}
  \hfill
  \subfloat[Inference on X-ray diffraction dataset]{\includegraphics[width=0.45\textwidth]{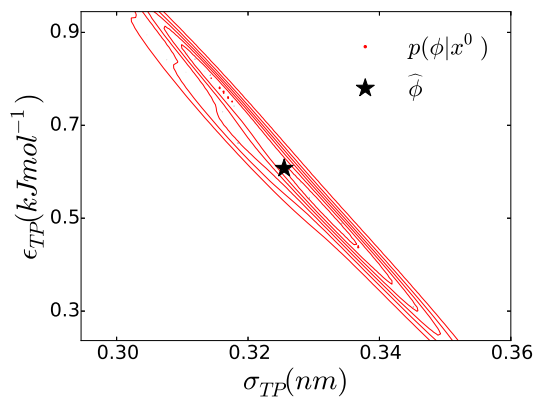}
  \label{fig:md_apmcabc_exp_2}}
    \caption{\textbf{X-ray diffraction dataset (Water):} \textbf{(a)} The radial distribution function of $O-O$ of water obtained by X-ray diffraction and reported in \cite{skinner2013benchmark}. \textbf{(b)} Contour plot of posterior distribution $p(\parametertheo|\dataObs)$ (red), Bayes estimate $\widehat{\parametertheo}\equiv (\hat{\sigma}_{TP},\hat{\epsilon}_{TP}) = 
  (0.326(nm), 0.607(kJ mol^{-1}))$ (black star). A strong negative posterior correlation of $-0.99$ between the parameters is present. The posterior distribution is obtained using a Gaussian kernel density estimator with bandwidth $0.7$.}
  \label{fig:xray_data}
\end{figure}

\paragraph*{X-ray Diffraction Dataset.}
Next we consider an experimental dataset where only radial distribution function of $O-O$ of water using the X-ray diffraction is obtained  \citep{skinner2013benchmark}. The experimentally obtained radial distribution function of $O-O$ is shown in Figure~\ref{fig:exprdfOO_2}, for further details we point readers to \cite{skinner2013benchmark}. Additionally, we use the value of the self-diffusion coefficient, $1.3e^{-5} cm^2 s^{-1}$, reported in \cite{vega2011simulating}. The experimental values of the other summary statistics calculated are:
$S_4^{e} = 0.98$, $S_5^{e}= 0.347(nm)$, $S_6^{e}= 0.141$ and $S_7^{e}= 2.57$, $S_8^{e}= 0.281(nm)$. In the absence of radial distribution function of $O-H$, we only consider these 5 summary statistics and the following distance function between them as the discrepancy measure:
\begin{eqnarray}
\label{eq:discrep_measure_partial}
\tilde{\distdata}_{TP}(\data^{(1)}, \data^{(2)})
&:=&\tilde{\distdata}_{TP}\left({\summaryfunc}_{TP}(\data^{(1)}), {\summaryfunc}_{TP}(\data^{(2)})\right)\\\nonumber
&=& \frac{1}{6}\sum_{i=4}^9|S_i^{(1)}-S_i^{(2)}|
\end{eqnarray}

In Figure~\ref{fig:md_apmcabc_exp_2}, we plot the posterior distribution of $\parametertheo$ and the Bayes estimates $\widehat{\parametertheo}\equiv (\hat{\sigma}_{TP},\hat{\epsilon}_{TP}) = (0.326(nm),  0.607(kJ mol^{-1}))$ obtained using the proposed Bayesian inference scheme. Also in this case, we see a strong negative posterior correlation of $-0.99$ between 
$\sigma_{TP}$ and $\epsilon_{TP}$.

\section{Model Prediction and Validation}
\label{sec:pred_uncertainty}
\subsubsection*{Model Prediction}
\begin{figure}[htbp]
  \centering
\includegraphics[width=0.5\textwidth]{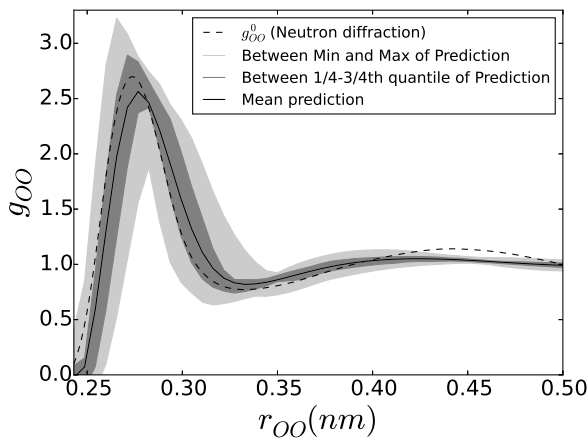}
    \caption{\textbf{Prediction of radial distribution function (Neutron diffraction):} Radial distribution function of $O-O$ obtained using Neutron diffraction (dashed), mean posterior prediction (solid) and the area between the minimum and maximum and the $\frac{1}{4}$-th and the $\frac{3}{4}$-th quantile of the posterior prediction using posterior samples reported in Figure~\ref{fig:md_apmcabc_exp}. We notice a close similarity between experimentally obtained and mean predicted data with reference to the properties of the first hydration shell.}
  \label{fig:pred_neutron_data}
\end{figure}

\begin{figure}[htbp]
  \centering
\includegraphics[width=0.5\textwidth]{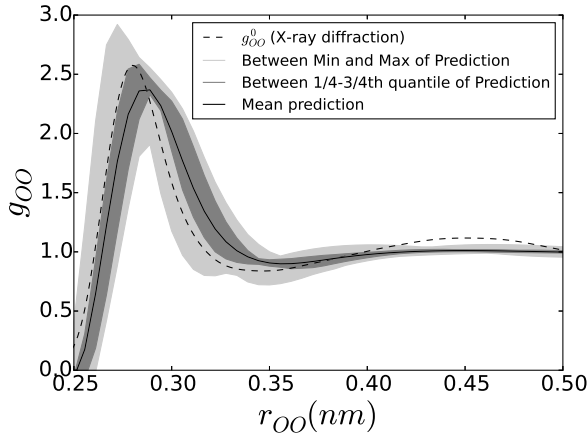}
    \caption{\textbf{Prediction of radial distribution function (X-ray diffraction):} Radial distribution function of $O-O$ obtained using X-ray diffraction (dashed), mean posterior prediction (solid) and the area between the minimum and maximum and the $\frac{1}{4}$-th and the $\frac{3}{4}$-th quantile of the posterior prediction using posterior samples reported in Figure~\ref{fig:md_apmcabc_exp_2}. We notice a close similarity between experimentally obtained and mean predicted data with reference to the properties of the first hydration shell.}
  \label{fig:pred_xray_data}
\end{figure}

As mentioned before, we can quantify the model prediction uncertainty, by simulating dataset from the MD simulation model using values of the parameters randomly drawn from the inferred posterior distribution. 
Using the posterior distribution inferred and illustrated in Figure~\ref{fig:md_apmcabc_exp} \& \ref{fig:md_apmcabc_exp_2}, we provide posterior prediction for the experimentally obtained radial distribution function of $O-O$ in Figure~\ref{fig:pred_neutron_data} \& \ref{fig:pred_xray_data}. We also illustrate the area between the minimum and maximum of the predicted datasets and the $\frac{1}{4}$-th and the $\frac{3}{4}$-th quantile of the predicted datasets with the light and dark gray color correspondingly, to assess the quality of the fit and the prediction uncertainty.  

Here we note that the experimental dataset is mostly within the prediction band (indeed, a large portion of values are inside the $\frac{1}{4}$-th and the $\frac{3}{4}$-th quantile of the predicted datasets), indicating a good predictive performance, except near the second maximum of $g_{OO}$. The mismatch near  $g_{OO}$, can be explained by the absence of the location and height of the the second maximum of $g_{OO}$ among the chosen summary statistics as reported in Section~\ref{sec:abcdetial}. This mismatch can be resolved by including those  summary statistics and modifying the discrepancy measure accordingly, which illustrates the flexibility of ABC algorithms. 

\subsubsection*{Validation}
For further validation of our inferred TIP4P force-fields from experimental dataset (X-ray and Neutron diffraction), in Table~\ref{table:validation_water} we compare values of a set of properties, which have not been used for parameterization. 
The properties considered here for comparison are, the heat capacity ($Cp \ cal mol^{-1} K^{-1}$) and the density ($\rho\ g cm^{-3}$) of liquid water at $298 K$ and of ice at $250 K$, as well as the isothermal compressibility ($\kappa_T\ 10^{-6} /bar$) and the dielectric constant ($\xi$) of water at $298 K$. 
The values of these properties were simulated (by performing $5\ ns$ NPT simulations at $298 K$ and $250 K$ using the forward simulation models described in Section~\ref{sec:forcefields}), using estimated 
$\widehat{\parametertheo}\equiv(\widehat{\sigma}_{TP},\widehat{\epsilon}_{TP})$ from both the dataset and the original TIP4P parameterization $\parametertheo^0\equiv (\sigma_{TP}^0,\epsilon_{TP}^0)$ and then compared with their experimental values obtained  under atmospheric pressure. 
\begin{table}[htbp]
\begin{minipage}{\textwidth}
\caption{\textbf{Comparison of properties (TIP4P Water)}: Values of heat capacity ($Cp$), density ($\rho$), isothermal compressibility ($\kappa_T$) and the dielectric constant ($\xi$), experimentally obtained (Expt.) under atmospheric pressure and simulated using estimated $\widehat{\parametertheo}\equiv(\widehat{\sigma}_{TP},\widehat{\epsilon}_{TP})$ from the two datasets (Neutron diffraction and X-ray diffraction) and the original TIP4P parameterization $\parametertheo^0\equiv (\sigma_{TP}^0,\epsilon_{TP}^0)$. These properties were not used 
in the ABC inference procedure to calibrate/estimate the parameters.}
\label{table:validation_water}
\begin{center}
\begin{tabular}{l|l|c|c|c|c}
&Prop. & Expt. & TIP4P & Neutron diff. & X-ray diff.  \\\hline
\multirow{ 2}{*}{Ice ($250K$)} &$Cp$ & 8.3 & 14.7 &
12.47  & 20.02 \\
&$\rho$ & 0.92  & 0.937 & 0.913 & 1 \\\hline
\multirow{ 5}{*}{Water ($298K$)}
&$Cp$ & 18  & 20 & 20.1 & 18.3 \\
&$\rho$ & 0.997 & 0.988 & 0.958 & 0.854  \\
&$\kappa_T$& 45.3  & 59 & 57.5 & 79.1 \\
&$\xi$ & 78.5  & 50 & 47 & 43 \\\hline
\end{tabular}
\end{center}
\end{minipage} 
\end{table}

The simulated values of the properties obtained using the estimated parametrization from Neutron diffraction dataset are closer to the experimental values than the ones simulated using the estimated parametrization from X-ray diffraction dataset, in all cases but for the heat capacity of the liquid water. We notice that the X-ray diffraction dataset did not have radial distribution function of $O-H$, and the predictive performance of the inference scheme suffers because of the absence of this summary statistics. 
Remarkably, the parametrization based on Neutron diffraction dataset is able to better predict the values of heat capacity ($Cp$) and density ($\rho$) of ice and the isothermal compressibility ($\kappa_T$) of water than the normal TIP4P parametrization. However, it predicts worse the dielectric constant ($\xi$) and the density ($\rho$) of water at $298K$. 

\section{Discussion}
We propose a Bayesian inference framework to calibrate the force-field parameters of molecular  systems, without a Gaussian model for the system uncertainty. 
This is achieved using approximate Bayesian computation, specifically with the help of adaptive population Monte Carlo ABC algorithm and High Performance Computing. The imbalance in the simulation time of MD simulations for different parameter values was resolved using a dynamic-allocation scheme for MPI. 
The methodology is illustrated by learning the force-field parameters of Lennard-Jones potential of helium and TIP4P system of water, both for simulated and experimental datasets. 
The proposed methodology has faster convergence and better scale-up performance compared to the ABCsubsim algorithm previously used to calibrate force-fields, thus contributing to a more efficient uncertainty quantification and parameter estimation of the force-fields.

Bayesian inference provides posterior distribution of the parameters, hence we are able to compute their posterior correlation  in a data-driven manner, which turns out to be strongly negative. Further, the negative correlation structure between non-bonded force-fields has been observed across two different experimental datasets, obtained using Neutron and X-ray diffraction. This points towards a strong underlying mechanism. Though further investigation is necessary to understand this mechanism, we 
hypothesize that:
At smaller values of $\sigma$, the repulsion is present at 
shorter distances and oxygen atoms can come closer. For oxygen atoms to come closer, while maintaining the same summary statistics (i.e the same number of first neighbors), without creating vacuum vacancies, there should be more attraction between each other. This scenario is equivalent to a bigger $\epsilon$. The opposite holds for higher values of $\sigma$.
Further using the posterior distribution, we were also able to quantify the model prediction uncertainty, which gives us an assessment on how good the model fit and prediction uncertainty are. 

We foresee that, for the calibration of challenging force-fields, more sophisticated ABC algorithms might be needed, possibly using surrogate and less computationally intensive models \citep{meeds2014gps}. 
Additionally, ABC model selection, without the need of Gaussian modeling of the uncertainty, can be used to learn the most suitable force-field formalism for an experimental dataset. This will be a line of future investigation.
Finally, in more complex systems and/or in a more complicated parameter space, reaching ergodicity is often challenging due to high free energy barriers, hampering a fast exploration of the free energy landscape, hence biasing the summary statistics. In the future we foresee combining our APMCABC framework to state of the art enhanced sampling methods such as Metadynamics, Umbrella sampling and Transition Path Sampling          \citep{Laio2002,Torrie1977,Brotzakis2016a}.

\section{Acknowledgments}
We thank Alessandro Laio and Michele 
Parrinello for suggesting the specific force-field of TIP4P,  
for insightful comments on the choice of the summary statistics and for discussing earlier drafts of this manuscript. 
This research was supported by Swiss National Science Foundation Grant No. 105218\_163196 (Statistical Inference on Large-Scale Mechanistic Network Models). We also thank Marcel Schoengens, CSCS, ETH Zurich for help regarding HPC and the Swiss National Super Computing Center for providing computing resources.\\\newline

\bibliographystyle{unsrtnat}


\end{document}